\documentstyle[multicol,aps,prl,psfig]{revtex} 

\renewcommand{\narrowtext}{\begin{multicols}{2} \global\columnwidth20.5pc}
\renewcommand{\widetext}{\end{multicols} \global\columnwidth42.5pc}
\def\top#1{\vskip #1\begin{picture}(290,80)(80,500)\thinlines \put(
65,500){\line( 1, 0){255}}\put(320,500){\line( 0, 1){
5}}\end{picture}}
\def\bottom#1{\vskip #1\begin{picture}(290,80)(80,500)\thinlines \put(
330,500){\line( 1, 0){255}}\put(330,500){\line( 0, -1){
5}}\end{picture}}

\def\al{\alpha}
\def\be{\beta}
\def\ga{\gamma}
\def\de{\delta}
\def\ep{\epsilon}

\def\ze{\zeta}
\def\et{\eta}

\def\la{\lambda}

\def\rh{\rho}

\def\si{\sigma}

\def\ph{\phi}

\def\ch{\chi}
\def\ps{\psi}

\def\Ga{\Gamma}

\def\Th{\Theta}

\def\Ps{\Psi}

\def\cl{{\cal L}}

\def\cO{{\cal O}}
\def\ap{\al^\prime}
\def\fr#1#2{{{#1} \over {#2}}}
\def\frac#1#2{\textstyle{{{#1} \over {#2}}}}
\def\pt#1{\phantom{#1}}
\def\prt{\partial}
\def\vev#1{\langle {#1}\rangle}
\def\ket#1{|{#1}\rangle}

\def\half{{\textstyle{1\over 2}}}
\def\lsim{\mathrel{\rlap{\lower4pt\hbox{\hskip1pt$\sim$}}
    \raise1pt\hbox{$<$}}}
\def\gsim{\mathrel{\rlap{\lower4pt\hbox{\hskip1pt$\sim$}}
    \raise1pt\hbox{$>$}}}

\newcommand{\beq}{\begin{equation}}
\newcommand{\eeq}{\end{equation}}
\newcommand{\bea}{\begin{eqnarray}}
\newcommand{\eea}{\end{eqnarray}}
\newcommand{\rf}[1]{(\ref{#1})}

\begin{document}

\title{ Stability, Causality, and Lorentz and CPT Violation }    
\author{V.\ Alan Kosteleck\'y and Ralf Lehnert}
\address{Physics Department, Indiana University, 
          Bloomington, IN 47405, U.S.A.}
\date{IUHET 427, October 2000; accepted in Physical Review D} 
\maketitle

\begin{abstract}
Stability and causality are investigated 
for quantum field theories incorporating Lorentz and CPT violation.
Explicit calculations in the quadratic sector of 
a general renormalizable lagrangian for a massive fermion
reveal that no difficulty arises for low energies
if the parameters controlling the breaking are small,
but for high energies 
either energy positivity or microcausality is violated
in some observer frame.
However,
this can be avoided if the lagrangian is the sub-Planck limit 
of a nonlocal theory with spontaneous Lorentz and CPT violation.
Our analysis supports the stability and causality of the
Lorentz- and CPT-violating standard-model extension
that would emerge at low energies from spontaneous breaking 
in a realistic string theory.

\end{abstract}


\narrowtext

\section{Introduction}
\label{intro}

Common folklore holds that the low-energy limit 
of any fundamental theory at the Planck scale
is necessarily a local relativistic quantum field theory.
If so,
this would make it difficult to identify experiments
showing directly any structural deviations from usual field theory
occurring at the Planck scale,
such as might perhaps be expected in string theories.
However,
this folklore is invalid 
if the fundamental theory violates one or more
of the basic tenets of relativistic field theories.
Remnant effects from the Planck scale
might then be detectable at low energies,
thereby providing valuable experimental information about
nature at the smallest scales.

Lorentz symmetry, stability, and causality are examples of
features normally expected to hold in physical quantum field theories.
In relativistic field theories,
stability and causality are closely intertwined with Lorentz invariance.
For example,
stability includes the need for energy positivity
of Fock states of arbitrary momenta,
while causality is implemented microscopically 
by the requirement that observables commute at spacelike separations
\cite{wp}.
Moreover,
both energy positivity and microcausality
are expected to hold in all observer inertial frames.

Although Lorentz symmetry is well established experimentally,
it lacks the essential status of stability and causality.
It would be difficult to make meaningful experimental predictions
in a theory without either stability or causality,
but a stable and causal theory without Lorentz symmetry 
could in principle still be acceptable. 
It is therefore worthwhile to consider the possibility
that Lorentz symmetry might be violated
and to examine the extent to which this violation 
conflicts with other fundamental properties of field theory.
In particular,
it would be of interest to establish the existence 
of a class of theories that incorporate Lorentz violation
but that nonetheless maintain both stability and causality.

Lorentz symmetry is also one of the key ingredients in the CPT theorem
\cite{cpt}.
This states under certain technical conditions
that CPT is an exact symmetry of local 
relativistic quantum field theories.
It is therefore to be expected that
investigations of theories with Lorentz violation
include a subset of cases in which CPT is also broken.

The present work is motivated by the development
over the past decade of
a framework allowing for Lorentz and CPT violation
within realistic models.
The basic idea is that spontaneous Lorentz violation
could occur in an underlying Lorentz-covariant theory
at the Planck scale
\cite{str}.
Under certain circumstances,
this would be accompanied by CPT violation.
This mechanism appears theoretically viable
and is motivated in part
by the demonstration that spontaneous Lorentz and CPT violation
can occur in the context of string theories
with otherwise Lorentz-covariant dynamics.
Lorentz- and CPT-violating effects could therefore
provide a unique low-energy signature for 
qualitatively new physics from the Planck scale.

At presently accessible energy scales,
these ideas lead to a phenomenology for Lorentz and CPT violation
at the level of the standard model and quantum electrodynamics (QED)
\cite{kp}.
A general standard-model extension has been developed 
that provides a quantitative microscopic framework
for Lorentz and CPT violation 
\cite{ck}.
It preserves the usual SU(3)$\times$SU(2)$\times$U(1) gauge structure
and is power-counting renormalizable.
Energy and momentum are conserved,
and conventional canonical methods for quantization apply.
The origin of the Lorentz violation in spontaneous symmetry breaking
implies that the standard-model extension
is covariant under observer Lorentz transformations:
rotations or boosts of an observer's inertial frame
leave the physics unaffected.
The apparent Lorentz violations in the theory are associated
with particle Lorentz transformations,
which are rotations or boosts of the localized fields in a fixed observer
inertial frame.

Since the standard-model extension is formulated at the level
of the known elementary particles,
it provides a quantitative basis on which to analyze 
a wide variety of Lorentz and CPT tests.
In the QED context,
investigations to date include
tests in Penning traps
\cite{bkr,gg,rm,hd},
studies of photon birefringence and radiative effects
\cite{ck,cfj,jk},
clock-comparison tests
\cite{cl,lh,rs,db,rw},
experiments with spin-polarized matter
\cite{bk,ah},
hydrogen and antihydrogen spectroscopy
\cite{bkr2,dp},
and studies of muons
\cite{bkl,vh}.
In the broader context of the standard-model extension,
studies of neutral-meson systems
\cite{mesons,kexpt,bexpt},
baryogenesis
\cite{bckp},
cosmic rays
\cite{cg,bc},
and neutrinos
\cite{ck,cg,sp}
have been performed.
Present experimental sensitivities are 
sufficient to detect Planck-suppressed effects.
Moreover,
the next generation of tests is expected to improve these results,
in some cases by one or more orders of magnitude. 

Given the substantial progress on the experimental front,
it is of interest to study the regime of validity within which 
the standard-model extension can be applied directly 
and to develop a methodology for handling the corrections 
that are expected at high energies.
Initiating this program is one of the goals of the present work.
The point is that the standard-model extension 
contains the low-energy limit of any realistic fundamental theory 
incorporating spontaneous Lorentz and CPT violation,
and on general grounds it is expected to have a range of validity 
comparable to that of the standard model at sub-Planck energies.
However,
as Planck energies are approached,
nonrenormalizable operators negligible at low energies
should acquire importance.
Since stability and causality are deeply related to
Lorentz symmetry at the level of renormalizable quantum field theory,
imposing them as requirements 
at high scales in the context of the standard-model extension
might be expected to yield interesting insights
into the structure of the nonrenormalizable terms.

The present work contains an investigation of
the role of stability and causality 
in Lorentz- and CPT-violating theories,
with particular emphasis on notions relevant
to the fermion sector of the standard-model extension.
We approach the subject by studying  
the quadratic fermion part of 
a general renormalizable lagrangian 
with explicit Lorentz- and CPT-breaking terms.
It is the single-fermion limit of the free-matter sector
in the general standard-model extension.
As a necessary part of the analysis,
we develop further the results of Ref.\ \cite{ck}
on the relativistic quantum mechanics of this theory
and perform the corresponding free-field quantization.
These results provide a complete quantization of the free-fermion sector
of the Lorentz- and CPT-violating QED extension,
including details such as the explicit general form 
of the one-particle dispersion relation.
Interactions can be handled in the usual perturbative manner
\cite{ck}.

One of our goals is to establish the nature of the difficulties
facing theories with explicit Lorentz violation,
however small.
We find violations of stability or causality
occur for momenta outside a scale determined 
by the size of the explicit breaking terms.
Although the scale in question may be large,
consistency problems are typically present for
any conventional quantum field theory of fermions 
with explicit Lorentz violation
\cite{ha}.

Another goal is to understand the mechanism
by which spontaneous Lorentz breaking in string theory
could overcome these difficulties.
By itself,
spontaneous Lorentz violation is an important ingredient.
However,
avoiding the problems with stability and causality
seems to require in addition its transcendental suppression 
at high energies in the one-particle dispersion relations,
through the appearance of nonrenormalizable terms
that are unimportant at low energies.
Interestingly,
this requirement naturally leads to field interactions of a type
related to those found in string field theory.

The analysis in this work leaves unaddressed several  
interesting theoretical issues associated with the transition 
from a fundamental theory with spontaneous Lorentz and CPT violation
at the Planck scale to the standard-model extension.
These include the development of the observed hierarchy
of scales in nature,
the role of fluctuations about the tensor expectation values
generating the extra terms in the standard-model extension,
the explicit incorporation of gravity,
and implications of nonminimality in the usual standard model
such as supersymmetry and gauge-group unification.
Although important in the development of a complete understanding,
these issues lie beyond the present scope.

The results developed in this work provide both 
a guide to the regime of validity of theories 
with explicit Lorentz violation 
and insight into the nature of the expected nonrenormalizable corrections 
to the standard-model extension 
emerging as the Planck scale is approached.
The twin demands of stability and causality 
lead from a renormalizable field theory
to a nonlocal theory incorporating spontaneous Lorentz breaking.
This supports the idea that the experimental observation 
of Lorentz violation would provide unique evidence 
for the nonlocality of nature at the Planck scale.

The outline of this paper is as follows.
Some basics are provided in section II.
Section III studies
relativistic quantum mechanics in a class of convenient inertial frames.
Section IV performs the canonical quantization of the field theory
and investigates stability and causality in arbitrary frames.
The issue of how the associated problems are resolved 
in the context of spontaneous Lorentz and CPT breaking 
in a fundamental theory is discussed in section V.
Finally,
a summary is provided in section VI.
Throughout,
we adopt the notations and conventions
of Ref.\ \cite{ck}.

\section{Some basics}

In this section,
we provide background material
and introduce some basic information
used in later sections of this work.
Some of this material is discussed in more detail
in Ref.\ \cite{ck}.

A general form for the quadratic sector of a renormalizable 
Lorentz- and CPT-violating lagrangian
describing a single massive spin-$\half$ Dirac fermion
is \cite{ck}:
\beq
\cl = \half i\overline{\psi} \Ga^\nu
\stackrel{\leftrightarrow}
{\prt}_{\nu}
\hspace{-.1cm}{\psi}
-\overline{\psi}M{\psi},
\label{lagr}
\eeq
where
\bea
{\Ga}^{\nu}&:=&{\ga}^{\nu}+c^{\mu \nu}
{\ga}_{\mu}+d^{\mu \nu}{\ga}_{5} {\ga}_{\mu}
\nonumber\\
&&\qquad
+e^{\nu}+if^{\nu}{\ga}_{5}
+\frac{1}{2}g^{\la \mu \nu}
{\sigma}_{\la \mu}
\label{Gam}
\eea
and
\beq
M:=m+a_{\mu}{\ga}^{\mu}+b_{\mu}{\ga}_{5}
{\ga}^{\mu}+\frac{1}{2}H^{\mu \nu}
{\sigma}_{\mu \nu}.
\label{M}
\eeq
In the above equations,
the gamma matrices $1$, $\ga_5$, $\ga^{\mu}$,
$\ga_5\ga^{\mu}$, $\sigma^{\mu \nu}$
have conventional properties.
In the context of the standard-model and QED extensions,
the parameters
$a_{\mu}$, $b_{\mu}$, $c_{\mu\nu}$, $\ldots$, $H_{\mu \nu}$
are determined by expectation values of Lorentz tensors 
arising from spontaneous Lorentz breaking
in a more fundamental theory.

For definiteness,
it is assumed throughout this work that the mass $m$
of the fermion is nonzero.
Our methods can in many cases be directly extended 
to the massless situation,
although the distinctions between finite- and zero-mass
representations of the Lorentz group
introduce some additional complications
that lie beyond our present scope.
In any case,
for most applications in the context of
the fermionic sector of the standard-model extension,
a nonzero mass is appropriate.
One possible exception is the study of neutrinos,
including neutrino oscillations.
If neutrinos have mass then the results below can be applied,
with minor modifications for Majorana fermions as necessary.
If one or more neutrinos are massless, then more care may be required.

Hermiticity of the lagrangian \rf{lagr}
implies that the coefficients for Lorentz violation are all real.
Moreover,
$c_{\mu \nu}$ and $d_{\mu \nu}$ can be taken as traceless,
$g_{\la \mu \nu}$ antisymmetric in its first two indices,
and $H_{\mu \nu}$ antisymmetric.
All the parameters violate particle Lorentz invariance,
while $a_{\mu}$, $b_{\mu}$, $e_{\mu}$, $f_{\mu}$, $g_{\la\mu\nu}$
also break CPT.
The coefficients in Eq.\ \rf{Gam} are dimensionless,
while those in Eq.\ \rf{M} have dimensions of mass.
The reader is warned that field redefinitions may eliminate
some of these coefficients without altering the physics
\cite{ck}.
For example,
introducing a nonzero coefficient $a_\mu$ 
in a single-fermion theory such as \rf{lagr}
has no observable consequences.
However,
$a_\mu$-type coefficients can lead to physical effects
in more general multifermion theories,
including the standard-model extension.
For completeness,
we explicitly keep all terms in Eq.\ \rf{lagr} in the present work.

The lagrangian \rf{lagr}
is independent of the coordinate system.
Observations made by any two inertial observers
can be related by coordinate transformations,
called observer Lorentz transformations.
Since Eq.\ \rf{lagr} is a scalar under these transformations,
the theory exhibits observer Lorentz symmetry.
However,
in Eq.\ \rf{lagr}
observer coordinate transformations differ profoundly
from boosts and rotations of particles or localized fields
within a fixed inertial frame.
The latter transformations,
called particle Lorentz transformations,
leave invariant the coefficients
$a_{\mu}$, $b_{\mu}$, $\ldots$, $H_{\mu \nu}$
and so can modify the physics
\cite{fn0}.
The particle Lorentz symmetry is therefore broken.

At the level of the present discussion,
the observer Lorentz symmetry of the theory \rf{lagr}
is a consequence of choosing a lagrangian 
invariant under Lorentz coordinate transformations.
More general classes of theories with explicit Lorentz violation
could in principle be considered.
For example,
the lagrangian might be taken to transform nontrivially
under the observer Lorentz group,
or perhaps as a scalar under 
some non-Lorentz coordinate transformation.
However,
these possibilities represent radical departures 
from conventional physics and lack motivation.
In contrast,
the explicit Lorentz-violating terms in the lagrangian \rf{lagr} 
could arise from a more fundamental theory with a lagrangian 
invariant under both observer and particle Lorentz symmetry,
provided the interactions in the theory are such as to cause
spontaneous Lorentz breaking.
If so,
then the coefficients 
$a_{\mu}$, $b_{\mu}$, $\ldots$, $H_{\mu \nu}$
for Lorentz and CPT violation are related 
to vacuum expectation values of Lorentz tensor fields
in the underlying theory,
and Eq.\ \rf{lagr} becomes a low-energy approximation
to this theory in the Lorentz-breaking vacuum.
The lagrangian \rf{lagr}
therefore serves as a single-fermion model for
the potentially realistic situation 
in which the standard-model extension emerges 
as the low-energy limit of spontaneous Lorentz violation 
in a fundamental theory at the Planck scale.

The distinction between observer and particle Lorentz transformations
implies a dual role for Lorentz symmetry 
in studying stability and causality of Eq.\ \rf{lagr}.
Thus,
if a theory is to be stable and causal,
then in a specified observer frame
the implications of energy positivity and microcausality
should hold for fields of different momenta
related through particle Lorentz transformations, 
while energy positivity and microcausality should hold 
in arbitrary inertial frames related by
observer Lorentz transformations.
In later sections,
it emerges that these two roles can be distinct.
For example,
a theory with spacelike 4-momentum for some one-particle states 
may maintain energy positivity 
under particle Lorentz transformations in a fixed frame,
but it will violate this requirement in certain other frames
obtained by suitable observer Lorentz transformations.

Since the various coefficients for Lorentz violation
in Eq.\ \rf{lagr} carry Minkowski indices,
they vary with the observer 
as appropriate representations of the noncompact Lorentz group SO(3,1) 
and are in this sense unbounded.
For some purposes,
it is useful to introduce a special class of inertial frames
in which the coefficients for Lorentz and CPT violation
represent only a small perturbation relative to the ordinary Dirac case.
We call a member of this class of frames 
a \it concordant frame. \rm
If Lorentz and CPT violation does indeed occur in nature,
then on experimental grounds
it must be true that 
any inertial frame in which the Earth moves nonrelativistically
can serve as a concordant frame.
The point is that 
no departures from Lorentz and CPT symmetry have been observed to date,
so any Lorentz and CPT violation 
in an Earth-based laboratory must be minuscule,
with the coefficients appearing in Eq.\ \rf{Gam} much smaller than 1
and those in Eq.\ \rf{M} much smaller than $m$.

In the present scenario,
the Lorentz- and CPT-violating effects are regarded as originating 
in a more fundamental theory at some large scale $M_P$.
It is plausible that $M_P$ is the Planck scale,
since this is the natural scale for an underlying theory
including gravity,
and in what follows we refer to it as such.
In any case,
it is expected that
observable effects in a low-energy theory with scale $m$
that arise from a fundamental theory with scale $M_P$ 
would be suppressed by some power of the ratio $m/M_P$.
It is therefore likely that the order of magnitude
of the coefficients appearing in Eq.\ \rf{Gam} 
is no greater than $m/M_P$,
while that of the coefficients in Eq.\ \rf{M} 
is no greater than $m^2/M_P$.

In conventional special relativity,
all inertial frames are equivalent
in the sense that high-energy physics in one frame
is in one-to-one correspondence with high-energy physics in any other frame.
However,
this equivalence fails in the present context.
The coefficients for Lorentz and CPT violation experienced
by a high-energy particle in one frame 
can differ substantially from those experienced 
by a high-energy particle in a second frame
because the particle Lorentz symmetry is broken.
In particular, 
this means that statements restricting attention to
Lorentz- and CPT-violating effects at high energies
may be observer dependent.

Given this ambiguity in the conventional notion of high energy,
it is useful to introduce a more precise definition.
For purposes of the present work,
the terminology of high and low energies 
relative to the scale of the underlying theory
is always taken to refer to a concordant frame
as defined above.
From an experimental point of view,
this terminology is sensible because 
by observation a laboratory frame moves nonrelativistically
with respect to a concordant frame.
The physics of high energies 
is therefore similar in both frames.

\section{Relativistic Quantum Mechanics}

In this section,
we study the lagrangian (\ref{lagr})
in the context of relativistic quantum mechanics.
The corresponding hermitian hamiltonian is derived,
and the associated dispersion relation is obtained.
We discuss properties of the eigenspinors
and determine the general solution of the equations of motion.
Throughout this section,
we work exclusively in a concordant frame
as defined in section II.

\subsection{Hamiltonian}

The construction of the relativistic quantum hamiltonian $H$
from the lagrangian $\cl$ of Eq.\ \rf{lagr}
requires care because $\cl$ contains time-derivative terms
in addition to the usual one.
In the concordant frame and a large class of associated observer frames,
this difficulty can be resolved by a spinor redefinition
chosen to eliminate the time-derivative couplings
\cite{bkr}.
Writing $\psi =A\chi$,
we require the non-singular matrix $A$
to be spacetime independent and to satisfy
\beq
A^{\dagger}\ga^0\Ga^0A=I ,
\label{redef}
\eeq
where $I$
is the $4\times4$ unit matrix.
With this choice,
$\cal L[\chi]$ 
contains no time derivatives outside the usual term 
$\frac{1}{2}{\it i}\overline{\chi}
{\ga}^0\hspace{-.15cm}
\stackrel{\leftrightarrow}
{\prt}_0\hspace{-.1cm}{\chi}$.
This spinor redefinition amounts to a change of basis in spinor space, 
and as such it leaves unchanged the physics.
Note that its explicit form depends on the choice of inertial frame.

It can be shown that $A$ exists if and only if 
all the eigenvalues of $\ga^0\Ga^0$ are positive.
First,
recall that an equivalence relation 
of the form $A^\dagger X A = Y$
between hermitian matrices $X$, $Y$
is called a congruence
\cite{lt}.
In the present case,
since both $I$ and $\ga^0\Ga^0$ are hermitian,
$A$ exists if and only if  
$\ga^0\Ga^0$ is congruent to $I$.
Next,
recall Sylvester's law of inertia,
which implies that under a congruence 
the number of positive eigenvalues
of a hermitian matrix is invariant.
Since $I$ has all positive eigenvalues,
the claimed result holds.

It follows that $A$ always exists in the concordant frame.
Define a matrix $\ep^0$ such that the zero component of 
Eq.\ \rf{Gam} can be written in the form
$\Ga^0=\ga^0(I+\ep^0)$.
Since the components of $\ep^0$
are small compared to 1 
in the concordant frame by definition,
the eigenvalues of $\ga^0\Ga^0=I+\ep^0$ are indeed positive
and $A$ therefore exists.

In Appendix A,
we obtain an upper bound on the size of the
coefficients for Lorentz and CPT breaking such that $A$ can exist.
The bound is expressed in terms of a quantity $\de^0$,
defined as the largest absolute value of 
certain coefficients for Lorentz and CPT violation:
\beq
\de^0=\max_{\mu\nu} 
\{|c_{\mu 0}|, |d_{\mu 0}|, |e_0|, |f_0|, |g_{\mu \nu 0}|\} .
\label{defd}
\eeq
We prove that $\de^0<1/480$
suffices for the spinor redefinition to exist.
The numerical value of this bound is far larger than the 
maximum size of $\de^0$ likely to be allowed on experimental grounds,
showing that the spinor redefinition 
indeed exists for the realistic situation.
Although it is sufficient for our purposes,
this bound is not sharp.
A determination of the sharp bound would be of interest.
We conjecture it is of order 1.

Once the spinor redefinition has been performed,
the Euler-Lagrange equations generate
a modified Dirac equation in terms of the new spinor $\ch$.
It can be written as
\beq
(i{\prt}_0-H){\chi}=0 ,
\label{dirac}
\eeq
where the hamiltonian
\beq
H=A^{\dagger}\ga^0(i\Ga^j
{\prt}_j-M)A
\label{hamilton}
\eeq
is hermitian,
as desired.
Explicit forms for this hamiltonian can
be found in Ref.\ \cite{cl}.

\subsection{Dispersion relation}

As usual,
a solution to Eq.\ \rf{dirac} is 
a superposition of plane waves of the form
\beq
\ch (x)=e^{-i\la_{\mu}x^{\mu}}
w(\vec{\la}) .
\label{wave}
\eeq
Here,
the 4-spinor $w(\vec{\la})$ must obey
\beq
({\la}_0-H) w(\vec{\la})=0 ,
\label{spinor}
\eeq
where $H$ is now understood to be in $\la$-momentum space,
and ${\la}_{\mu}$ must satisfy
the dispersion relation
\beq
\det({\la}_0-H) = 0 .
\label{disp1}
\eeq
An alternative equivalent form for the dispersion relation is
\beq
\det(\Ga^\mu {\la}_{\mu}-M)
= 0 ,
\label{disp}
\eeq
since the non-singular matrices 
$\ga^0$, $A$ and $A^{\dagger}$
relating the two forms of the Dirac equations
contribute only overall multiplicative factors to the determinant.

To obtain an explicit expression
for the dispersion relation,
we write the matrix
$\Ga^{\mu}\la_{\mu}-M$ as
\bea
\Ga^{\mu}\la_{\mu}-M
&=&
S+P\ga_5+V^{\mu}\ga_{\mu}
\nonumber\\
&&\qquad
+A^{\mu}\ga_5\ga_{\mu}+T^{\mu\nu}\si_{\mu\nu} ,
\label{convenientway}
\eea
where we have introduced
\begin{eqnarray}
&S = e^{\mu}\la_{\mu}-m ,
\quad P = f^{\mu}\la_{\mu},&
\nonumber\\
&V^{\mu} = \la^{\mu}
+c^{\mu\nu}\la_{\nu}-a^{\mu} ,
\quad A^{\mu} = d^{\mu\nu}
\la_{\nu}-b^{\mu},&
\nonumber\\
&\quad T^{\mu\nu} = \frac{1}{2}g^{\mu\nu\rh}
\la_{\rh}-\frac{1}{2}H^{\mu\nu}.&
\label{abbrev}
\end{eqnarray}
Expansion of the determinant of this matrix yields
\bea
0&=&4(V_{\mu}A_{\nu}-A_{\mu}V_{\nu}
+V_{\mu}V_{\nu}+A_{\mu}A_{\nu}
\nonumber\\
&&\qquad
+PT_{\mu\nu}-S\tilde T_{\mu\nu}
+\frac{1}{2}T_{\mu\al}
T^{\al}_{\pt{\al}\nu}
-\frac{1}{2}T^2{\eta}_{\mu\nu})^2
\nonumber\\
&&\quad 
+(V^2-A^2-S^2-P^2)^2-4(V^2-A^2)^2
\nonumber\\
&&\quad
+6(\ep_{\mu\nu\al\be}
A^{\al}V^{\be})^2 ,
\label{convenientform}
\eea
where
$\tilde T^{\mu\nu}= \frac{1}{2} \ep^{\mu\nu\al\be} T_{\al\be}$
denotes the dual tensor.

The dispersion relation \rf{convenientform}
can be viewed as a quartic equation for
$\la^0(\vec{\la})$.
In principle,
it permits the explicit determination
of the exact eigenenergies of a particle
with given 3-momentum in the presence of Lorentz and CPT violation.
Various approximate solutions can also be obtained.
For example, 
in certain applications only the leading-order corrections
to the conventional eigenenergies are of interest.
However,
we caution the reader that these cannot necessarily
be obtained by keeping only leading contributions
to the coefficients of the momentum in the dispersion relation 
and solving for the energies,
as is argued in some of the published literature
\cite{fn1}.

Many of the relevant properties 
of the dispersion relation 
can be established without an explicit algebraic solution.
For example,
since $H$ is hermitian
all four roots of the dispersion relation must be real.
It follows from Eq.\ \rf{disp}
that the roots are independent of the spinor redefinition \rf{redef}, 
as expected.
This equation also implies that the dispersion relation
is observer Lorentz invariant and hence
that $\la_{\mu}$ must be an observer Lorentz 4-vector.

In general,
the fourfold degeneracy of the magnitudes of the roots
of Eq.\ \rf{disp} is lifted,
a feature different from the conventional Dirac case. 
Since the Lorentz and CPT violation is small in the concordant frame,
one still anticipates two positive roots
$\la^0_{+(\al)}(\vec{\la})$,
$\al=1,2$, 
and two negative roots
$\la^0_{-(\al)}(\vec{\la})$.
In Appendix B,
we obtain a bound on the size of the coefficients for 
Lorentz and CPT violation such that this anticipation is correct.
The bound is in terms of a quantity $\de$,
defined as 
\bea
\delta&=&\max_{\mu,\nu, j} \{
|a_{\mu}|,
|b_{\mu}|,
m |c_{\mu j}|,
m |d_{\mu j}|, 
\nonumber\\
&&\qquad \qquad \qquad
m |e_j|, 
m |f_j|,
m |g_{\mu \nu j}|,
|H_{\mu \nu}|
\} ,
\label{defdel}
\eea
where the Greek indices range from 0 to 3
and the Latin index ranges from 1 to 3,
as usual.
We find that for $\delta<m/124$
the dispersion relation 
has two positive and two negative solutions,
as usual.
This bound is independent of the spinor redefinition.
Its numerical value is much larger than
experimental observations are likely to allow,
showing that the presence of Lorentz and CPT violation in nature
would indeed leave unaffected the counting of 
positive- and negative-energy solutions.
Although more than adequate for our purposes,
this bound is not sharp,
and it would be of interest to determine the sharp bound. 
We anticipate it is of order 1. 

Another important feature of the dispersion relation 
is the correspondence
\bea
&&\la^0_{-(1,2)}(\vec{\la}
,a_{\mu}, d_{\mu \nu},
e_{\mu}, f_{\mu}, H_{\mu \nu})=
\nonumber\\
&&
\qquad
-\la^0_{+(2,1)}(-\vec{\la}
,-a_{\mu}, -d_{\mu \nu},
-e_{\mu}, -f_{\mu}, -H_{\mu \nu})
\label{symmet}
\eea
between
the positive and negative solutions.
In this equation,
we have displayed only the dependence 
on the coefficients for Lorentz and CPT violation that change sign,
and it is understood that the other coefficients are held constant.
The numbering of the roots is chosen to
agree with the results in Ref.\ \cite{ck}.
Equation \rf{symmet} can be regarded 
as a consequence of the identity 
$\det(\Ga^\mu {\la}_{\mu}-M)=
\det[C(\Ga^\mu {\la}_{\mu}-M)C^{-1}]$,
where $C$ is the usual charge-conjugation matrix.
This implies the invariance of
$\det(\Ga^\mu {\la}_{\mu}-M)$
under the transformation
\bea
&&\{\vec{\la}
,a_{\mu}, d_{\mu \nu},
e_{\mu}, f_{\mu}, H_{\mu \nu}\}
\rightarrow
\nonumber\\
&&
\qquad
\{-\vec{\la}
,-a_{\mu}, -d_{\mu \nu},
-e_{\mu}, -f_{\mu}, -H_{\mu \nu}\} 
\eea
and leads to the correspondence \rf{symmet}. 

\subsection{Eigenspinors}

The eigenfunctions corresponding to the two negative roots
$\la^0_{-(\al)}$
can be reinterpreted as positive-energy
reversed-momentum wave functions
in the usual way.
We define
\begin{eqnarray}
\chi^{(\al)}_+ &
= & \exp (-ip^{(\al)}_u\!\!\cdot\! x)
\: u^{(\al)}(\vec{p}),
\nonumber \\
\chi^{(\al)}_- &
= & \exp (+ip^{(\al)}_v\!\!\cdot\! x)
\: v^{(\al)}(\vec{p}),
\label{wavefcn}
\end{eqnarray}
where 
$u^{(\al)}(\vec{p})$ and
$v^{(\al)}(\vec{p})$
are momentum-space spinors
and the 4-momenta
are given by
\begin{eqnarray}
p^{(\al)}_u =
(E^{(\al)}_u,\vec{p}) , & \quad
E^{(\al)}_u(\vec{p})= &
\la^0_{+(\al)}(\vec{p}) ,
\nonumber\\
p^{(\al)}_v =
(E^{(\al)}_v,\vec{p}) , & \quad
E^{(\al)}_v(\vec{p})= &
-\la^0_{-(\al)}(-\vec{p}) .
\label{momenta}
\end{eqnarray}

The symmetry (\ref{symmet}) of the dispersion relation
determines a relationship between the two sets of energies.
We find
\bea
&&E^{(1,2)}_v(\vec{p},
a_{\mu}, d_{\mu \nu},
e_{\mu}, f_{\mu}, H_{\mu \nu}) =
\nonumber\\
&&
\qquad
E^{(2,1)}_u(\vec{p},
-a_{\mu}, -d_{\mu \nu},
-e_{\mu}, -f_{\mu}, -H_{\mu \nu}) .
\label{ensymmetry}
\eea
Similarly,
the spinors are related by
\bea
&&v^{(1,2)}(\vec{p},
a_{\mu}, d_{\mu \nu},
e_{\mu}, f_{\mu}, H_{\mu \nu}) =
\nonumber\\
&&
\qquad
u^{(2,1)c}(\vec{p},
-a_{\mu}, -d_{\mu \nu},
-e_{\mu}, -f_{\mu}, -H_{\mu \nu}) ,
\label{spsymmetry}
\eea
where the superscript $c$ denotes a charge-conjugate spinor
defined by $w^c= C{\overline w}^T$, as usual.

The spinors $u$ and $v$ are the eigenvectors
of the hermitian matrix $H$
and they therefore span the spinor space.
Orthogonality of the eigenspinors is automatic
for nondegenerate eigenenergies
and in any case can be imposed by choice.
The normalization of $u$ and $v$
is constrained by the requirement $(\chi^c)^c=\chi$
but is otherwise arbitrary. 
For definiteness,
we choose the conditions
\begin{eqnarray}
u^{(\al)\dagger}(\vec{p})
u^{(\al^{\prime})}(\vec{p}) & =
& \delta^{\al\al^{\prime}}
\fr{E^{(\al)}_u}{m} ,
\nonumber\\
v^{(\al)\dagger}(\vec{p})
v^{(\al^{\prime})}(\vec{p}) & =
& \delta^{\al\al^{\prime}}
\fr{E^{(\al)}_v}{m} ,
\nonumber\\
u^{(\al)\dagger}(\vec{p})
v^{(\al^{\prime})}(-\vec{p}) &
= & 0 .
\label{onc}
\end{eqnarray}
Note, 
however,
that the conventional generalization of the orthogonality relation
involving the Dirac-conjugate spinors
$\overline{u}$ and $\overline{v}$ 
fails in the present case.
Equation (\ref{onc})
implies the completeness relation
\widetext
\top{-2.8cm}
\hglue -1 cm
\begin{eqnarray}
\sum_{\al=1}^2
\left(
\fr{m}{E^{(\al)}_u(\vec{p})}
u^{(\al)}(\vec{p})
\otimes
u^{(\al)\dagger}(\vec{p})+
\fr{m}{E^{(\al)}_v(-\vec{p})}
v^{(\al)}(-\vec{p})
\otimes
v^{(\al)\dagger}(-\vec{p})
\right)
=I .
\label{complete}
\end{eqnarray}

With the above definitions,
the general solution to the modified Dirac equation \rf{dirac}
can be written as
\bea
\chi(x)=
\int
\fr{d^3p}{(2\pi)^3}
\sum_{\al=1}^2
\left(
\fr {m} {E^{(\al)}_u} b_{(\al)}(\vec{p})
\exp(-ip^{(\al)}_u\!\cdot x) u^{(\al)}(\vec{p})
+ \fr{m}{E^{(\al)}_v} d^*_{(\al)}(\vec{p})
\exp(+ip^{(\al)}_v\!\cdot x) v^{(\al)}(\vec{p})
\right) ,
\label{solution}
\eea
\bottom{-2.7cm}
\narrowtext
\noindent
where $b_{(\al)}(\vec{p})$
and $d^*_{(\al)}(\vec{p})$
are Fourier coefficients,
as usual.
For simplicity,
the dependence of the eigenenergies and eigenspinors
on the coefficients for Lorentz and CPT violation 
is suppressed in this equation.

\section{Quantum Field Theory}

In this section,
we perform canonical quantization in a concordant frame
by demanding energy positivity,
as usual.
We then study the issues of stability and causality
in arbitrary frames.

\subsection{Canonical Quantization and Energy Positivity}

In the usual case,
straightforward canonical quantization of a Dirac fermion is inadequate
because the theory is singular.
Appropriate quantization conditions can be found
either by requiring the positivity of the conserved energy
or, 
more formally,
by extending the Dirac-bracket procedure to anticommuting fields
\cite{gt}.
We adopt the former procedure here.

We promote the complex weights in the expansion \rf{solution}
to operators on a Fock space.
The spinor $\ch$ thereby becomes a quantum field,
as does the spinor $\ps$.
The two fields are related 
through the redefinition $\ps = A \ch$,
where $A$ is the same matrix discussed in the previous subsection.

We impose the following nonvanishing anticommutation relations:
\bea
\{ b_{(\al)}(\vec{p}),
b_{(\al^{\prime})}^{\dagger}
(\vec{p}^{\:\prime}) \}
& = & (2\pi)^3\fr{E^{(\al)}_u}{m}
\delta_{\al\al^{\prime}}
\delta(\vec{p}-\vec{p}^{\:\prime}) ,
\nonumber\\
\{ d_{(\al)}(\vec{p}),
d_{(\al^{\prime})}^{\dagger}
(\vec{p}^{\:\prime})\}
& = & (2\pi)^3\fr{E^{(\al)}_v}{m}
\delta_{\al\al^{\prime}}
\delta(\vec{p}-\vec{p}^{\:\prime}) .
\label{commut}
\eea
These can be used to reconstruct 
the equal-time anticommutators for the fields $\ch$:
\bea
\{\ch_j(t,\vec{x}),
\overline{\ch}_l(t,\vec{x}^{\:\prime})
\ga^0_{lk}\}
&=&\de_{jk}\de^3(\vec{x}-\vec{x}
^{\:\prime}) , 
\nonumber\\
\{\ch_j(t,\vec{x}),
\ch_k(t,\vec{x}^{\:\prime})\}
&=&\{\overline{\ch}_l(t,\vec{x})
\ga^0_{lj},
\overline{\ch}_m(t,\vec{x}^{\:\prime})
\ga^0_{mk}\} 
\nonumber\\
&=& 0 ,
\label{fieldcommchi}
\eea
where the spinor indices $j,k,l,m$ are displayed for clarity.

The above expressions permit the derivation of 
the equal-time anticommutators for the original fields $\psi$ as
\bea
\{\psi_j(t,\vec{x}),
\overline{\psi}_l(t,\vec{x}^{\:\prime})
\Ga^0_{lk}\}
&=&\delta_{jk}\delta^3(\vec{x}-\vec{x}
^{\:\prime}) , 
\nonumber\\
\{\psi_j(t,\vec{x}),
\psi_k(t,\vec{x}^{\:\prime})\}
&=&\{\overline{\psi}_l(t,\vec{x})
\Ga^0_{lj},
\overline{\psi}_m(t,\vec{x}^{\:\prime})
\Ga^0_{mk}\} 
\nonumber\\
&=& 0 .
\label{fieldcomm}
\eea
Note that
$\pi_{\psi}= \overline{\psi}\Ga^0$
is the canonical conjugate of $\psi$,
paralleling the usual Dirac case.

The vacuum state $|0\rangle$ of the Hilbert space
in the concordant frame is defined by
\beq
b_{(\al)}(\vec{p})|0\rangle=0
,\qquad
d_{(\al)}(\vec{p})|0\rangle=0 .
\label{vacuum}
\eeq
The action of the creation operators
$b^{\dagger}_{(\al)}(\vec{p})$
and $d^{\dagger}_{(\al)}(\vec{p})$
on $|0\rangle$
produces states describing particles and antiparticles
with 4-momenta
$p^{(\al)}_u$ and $p^{(\al)}_v$, 
respectively.
This can be verified using the normal-ordered conserved momentum
\beq
P_{\mu}=\int d^3x~ :\Th_{\mu 0}:~,
\eeq
where
\beq
\Theta_{\mu\nu}=
\frac{1}{2}i\overline{\psi}
\Ga_{\mu}\stackrel{\leftrightarrow}
{\prt}_{\nu}\psi
\eeq
is the conserved canonical energy-momentum tensor.

In the present context,
the one-particle states carry the 4-momenta 
$p^{(\al)}_u$ and $p^{(\al)}_v$
introduced in the previous section.
It follows from Eq.\ \rf{momenta}
that the zero components of these 4-vectors 
are positive definite.
This validates the quantization ansatz \rf{commut}
in the concordant frame.

The lagrangian \rf{lagr} is observer Lorentz invariant by construction.
The observables resulting from quantization should therefore
be invariant or depend covariantly on the observer.
In the usual case,
Lorentz transformations are unitarily implemented
on the Hilbert space of states,
and so covariance follows directly.
In contrast,
in the present case
the coefficients for Lorentz and CPT violation carry spacetime indices,
and their values therefore depend on the observer.
This implies that the Fock spaces
constructed by different observers are inequivalent.
Nonetheless,
the invariance of observables may be implemented
by suitable mappings between the Fock spaces for any two observers.
These mappings then form a representation of the Lorentz group
with group multiplication being the mapping composition.
Note that the existence of this group structure is assured
if the Lorentz violation is spontaneous.
In this case,
although the observer Lorentz symmetry 
cannot be unitarily implemented on the Fock space,
the freedom to select the physical vacuum
among all Lorentz-equivalent choices
means that all observers have Fock spaces in one-to-one correspondence.

The field quantization presented above
can be performed provided
the bounds on $\de^0$ and $\de$ in section III
are satisfied,
so that the Lorentz-violating time-derivative terms
can be removed and the usual eigenenergy-sign structure holds. 
These conditions involve the size of individual
components of observer Lorentz tensors
and are thus inherently noninvariant
under observer Lorentz transformations.
There is therefore a class of observers,
strongly boosted relative to a concordant frame,
for whom these bounds are violated and  
the present technique of field quantization fails.
However,
as discussed above,
the observer Lorentz invariance guarantees
a one-to-one correspondence of the Fock spaces
among all observers,
so some difficulties must also exist 
even for the quantization scheme
in a concordant frame.
It turns out these are associated with the
stability and causality of the theory.
The next two subsections discuss 
these issues in detail.

\subsection{Stability}

In usual Lorentz-covariant free-field theories,
energy positivity in a particular frame
translates under certain assumptions
to the statement that the vacuum is stable in any frame.
One assumption is that 
the 4-momenta of all one-particle states in the particular frame
are timelike or lightlike with nonnegative 0th components.
This is satisfied in the usual Dirac theory.
Since an observer Lorentz transformation cannot change the sign 
of these 0th components,
energy positivity is in this case a Lorentz-invariant notion
even though it is a statement about a 4-vector component.

In the present case with Lorentz and CPT violation,
energy positivity in a concordant frame
is assured if the bound on $\de$ discussed in section IIIB
is satisfied.
However,
stability of the quantized theory in all observer frames
requires more than just energy positivity in a concordant frame.
In fact,
one of the usual assumptions fails:
some of the energy-momentum 4-vectors
solving the dispersion relation \rf{disp}
may under certain circumstances be spacelike
in all observer frames.

As an example,
consider the dispersion relation
\beq
(\la^2-b^2-m^2)^2+4b^2\la^2 -4(b\cdot\la)^2=0
\label{bdisp}
\eeq
for a model
with a $b_{\mu}$ coefficient only.
One can show that for any nonzero $b_\mu$,
no matter how small,
it is always possible to choose an observer frame
in which $b_{\mu}=(b_0,0,0,b_3)$
and ${b_3}^2>m^2+|b^{\mu}b_{\mu}|$.
Defining the real quantities $p_{\pm}$ by
\bea
{p_{\pm}}^2&=&(2{b_3}^2+b^2-m^2)
\nonumber\\
&& \qquad
\pm\sqrt{(2{b_3}^2+b^2-m^2)^2-(m^2+b^2)^2},
\label{moment}
\eea
the spacelike 4-vectors
${\la^{\mu}}_{\pm}=(0,0,0,p_{\pm})$
can be shown to satisfy the dispersion relation \rf{bdisp},
as the reader is invited to verify.
Moreover,
the existence of such spacelike solutions to the dispersion relation 
is unaffected by the inclusion of a nonzero $a_\mu$,
for example.

Although the instabilities introduced
by the existence of spacelike solutions exist in any frame, 
including a concordant frame as discussed below, 
they are most transparent by considering observer Lorentz boosts.
An appropriate observer boost involving a velocity less than 1
can always convert a spacelike vector with a positive 0th component
to one with a negative 0th component.
In the present instance,
this means that there exist otherwise acceptable observer frames
in which a single root of the dispersion relation 
involves both positive and negative energies.
In such frames,
the canonical quantization procedure fails.

In Figs.\ 1 and 2,
the appearance of negative energies in a strongly boosted frame
is illustrated for a model with only a nonzero $b_0$
in a concordant frame.
The dispersion relation 
as seen by an observer in a concordant frame
is shown in Fig.\ 1.
One of the two positive roots is displayed.
The energy is manifestly positive for all 3-momenta.
However,
the dispersion relation crosses the light cone
\cite{fn2}
at a finite value $\tilde M$ of the 3-momentum.
Beyond this value,
points lying on the curve can be regarded as represented
by spacelike vectors relative to the origin.
All these spacelike vectors have positive 0th components.

For a suitable boost,
some of the spacelike vectors are converted 
to spacelike vectors with negative 0th components.
Figure 2 shows the result of a large boost.
A portion of the dispersion relation has dipped below
the energy zero.
The corresponding negative-energy states represent a stability problem
for the theory when interactions are introduced.
We remark in passing that
under the same boost
the other roots of the dispersion relation are positioned 
so as to preclude eliminating the negative energies by a simple shift
of the energy zero.

\begin{figure}
\centerline{\psfig{figure=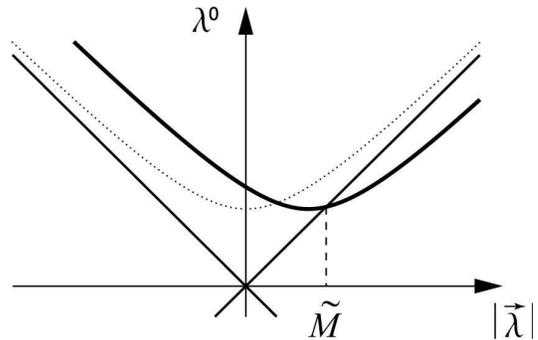,width=0.8\hsize}}
\smallskip
\caption{Dispersion relation for a model with only a 
large nonzero $b_0$ in a concordant frame.
One of the two positive roots is displayed.
It intersects the light cone at a 3-momentum
of magnitude $\tilde M$.
The dotted line is the conventional dispersion relation 
for a massive particle.}
\label{fig1}
\end{figure}

\begin{figure}
\centerline{\psfig{figure=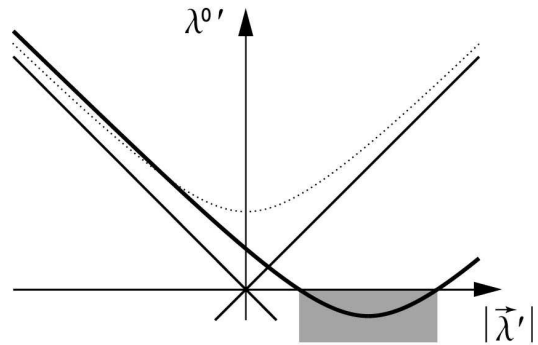,width=0.8\hsize}}
\smallskip
\caption{Dispersion relation for the model of Fig.\ 1
as seen by an observer strongly boosted 
relative to the concordant frame.
The occurrence of negative energies is apparent
in the shaded region.
The dotted line is the conventional dispersion relation 
for a massive particle.}
\label{fig2}
\end{figure}

The scale $\tilde{M}$ of the 3-momentum
at which the 4-momentum turns spacelike
can be calculated explicitly in various models.
For example,
consider the case of a timelike $b_{\mu}$,
as above.
In an observer frame with $b_{\mu}=(b_0,\vec{0})$,
we find
\bea
\tilde{M}&=&\fr{m^2+{b_0}^2}{2|b_0|}
\nonumber\\
& \gsim & {\cO}(M_P) .
\label{bscale}
\eea
The approximate equality in the last step
is attained for the case of a single suppression factor
from the Planck scale,
$b_0 \sim \cO(m^2/M_P)$,
following the discussion in section II.

This estimate reveals that the instabilities in the model
emerge only for Planck-scale 4-momenta in a concordant frame.
The corresponding negative energies 
appear only for observers undergoing a Planck-scale boost
relative to this frame.
It follows that 
the concordant-frame quantization we have presented above 
maintains stability for all experimentally attainable physical momenta
and in all experimentally attainable observer frames.

Inspection of the dispersion relation for the $b_\mu$ model
reveals that in all observer frames
the asymptotes of the dispersion relation 
are parallel to the usual light-cone asymptotes.
The behavior can also be seen in the example in Figs.\ 1 and 2.
We see that, 
to avoid spacelike 4-momenta, 
the asymptotes of the dispersion relation
must remain inside the usual light cone.
In terms of the group velocity $\vec v_g$ 
of a wave packet in the theory,
given as usual by
\beq
\vec v_g = \fr{\prt E}{\prt\vec{p}},
\label{vgr}
\eeq
this requirement on the asymptotes  
implies the following necessary condition for energy positivity: 
\beq
|\vec v_g | \ge 1,
\qquad 
|\vec{p}|\rightarrow\infty .
\label{cond1}
\eeq
The reader is reminded that 
the relation between momentum and group velocity is unconventional
\cite{ck}.
In particular, 
$\vec{p}$ and $\vec{v}_g$ need not be parallel.

Since the physics is invariant under observer boosts,
the appearance of negative energies in a strongly boosted frame
indicates that spacelike 4-momenta
lead to a stability problem also in a concordant frame,
albeit only for particles with energies exceeding the Planck scale.
As an illustration,
consider the following process in a concordant frame:
a Planck-energy fermion emits a virtual photon,
which then decays into a fermion-antifermion pair.
We can write this as
\beq
f_{+1}\longrightarrow f_{+1} +f_{+1}+\bar{f}_{-1} ,
\label{decay}
\eeq
where $f$ and $\bar{f}$ denote fermions and antifermions,
respectively,
and the subscript labels the helicity state.
In conventional QED,
this decay is kinematically forbidden
even though both the U(1) charge and angular momentum are conserved.
However,
for Planck energies it can occur 
in the context of the Lorentz- and CPT-violating QED extension
with a nonzero $b_0$ coefficient.
The dispersion relation for the 4-momentum $(E,\vec p)$ 
of a fermion of helicity $+1$ or an antifermion of helicity $-1$
is given in Appendix B of the first paper in Ref.\ \cite{ck} as
\beq
E=\sqrt{m^2+(|\vec{p}|-b_0)^2} .
\label{bOenergy}
\eeq
Taking for simplicity the 3-momentum $|\vec{q}|$
of the incoming fermion as 
\beq
|\vec{q}|=\fr{2m^2+{b_0}^2}{b_0}+b_0 \gsim {\cO}(M_P) ,
\label{pimag}
\eeq
we find the process \rf{decay} is kinematically allowed
with all final 3-momenta equal to $\vec{q}/3$.
A single-particle state describing a fermion
of sufficiently large 3-momentum \rf{pimag} and helicity +1 is 
therefore unstable.
The instability also occurs for other high-energy single-particle states,
although the final 3-momenta are then unequal.

It can be shown that an initial spacelike 4-momentum
is a necessary condition allowing the process \rf{decay},
as expected.
The decay process \rf{decay} could therefore occur repeatedly
in a cascade until the energy of the decay products
reaches the order of the Planck scale in a concordant frame.
Although unusual,
this behavior and related phenomena involving other decays  
might be phenomenologically admissible.
However,
in what follows we focus on the possibility of maintaining stability 
at the Planck scale despite the presence of Lorentz violation.

The conclusion that instabilities enter at ${\cO}(M_P)$,
as in Eq.\ \rf{bscale},
may fail for models with a nonzero coefficient $c_{\mu\nu}$.
This coefficient is special 
because the associated quadratic field term 
has the same general spinorial and derivative structure 
as the usual Dirac kinetic term,
and so it acts as a first-order correction
to an existing zeroth-order term.
No other Lorentz-violating term has this feature.

As an explicit example,
consider a model with only the coefficient $c_{00}$ nonzero 
in a concordant frame
\cite{fn3}.
The dispersion relation for this model in an arbitrary frame is
\beq
(\et_{\al\mu}+c_{\al\mu})
(\et_{\pt{\nu}\nu}^{\al}+c^\al_{\pt{\nu}\nu})
\la^\mu\la^\nu
-m^2 = 0.
\label{cdispcov}
\eeq
In the concordant frame,
this takes the form
\beq
\ze^2{\la_0}^2-{\vec{\la}}^2 -m^2=0 ,
\label{cdisp}
\eeq
where we define $\ze=1+c_{00}$.
For the case $c_{00}>0$,
we then find that spacelike 4-momenta occur at a scale $\tilde{M}$
given by
\bea
\tilde{M} & = & \fr{m}{\sqrt{\zeta^2-1}}
\approx \fr{1}{\sqrt{2c_{00}}}m+{\cO}(c_{00})
\nonumber\\
& \gsim & {\cO}(\sqrt{mM_P}) ,
\label{c00scale}
\eea
where in the last step the approximate equality
is attained for a single suppression factor from the Planck scale,
$c_{00} \sim \cO(m/M_P)$.

The result \rf{c00scale} implies that 
instabilities occur at energies well below
the scale $M_P$ of the underlying theory
in the $c_{00}$ model with $c_{00}>0$.
We show in the next section that if $c_{00}<0$ instead,
then microcausality violations arise at the same scale.
If these results continue to hold in the full underlying theory,
they could have observable physical implications.
As one example,
Coleman and Glashow have suggested 
\cite{cg} 
the interesting possibility
that high-energy effects from $c_{00}$-type terms 
might be responsible for the apparent excess of
cosmic rays in the region of $10^{19}$ GeV.
This scale is potentially comparable to $\sqrt{mM_P}$.
However,
if stability and causality are imposed on the theory, 
then the $c_{00}$ dispersion relation \rf{cdisp} must be modified.
This in turn is likely to modify the physical implications
at high energies.
In section V,
we discuss some possible high-energy corrections 
to Eq.\ \rf{cdisp} that would preserve stability and causality.
It would be of interest to revisit the cosmic-ray analysis 
in light of these requirements.

In any case,
given the impracticality of achieving Planck-scale 
energies or boosts in the laboratory,
the issues with spacelike 4-momenta are largely unimportant
at the level of the standard-model extension.
However,
they do confirm the expectation that corrections to the theory
at high energies are needed for complete stability.
Requiring stability therefore has the potential to provide insight
into the nature of the corrections.
This situation is qualitatively different from that occurring
in conventional special relativity,
where Planck-scale boosts are admissible 
without generating instabilities internal to the theory.
Since the standard-model extension contains 
all relevant renormalizable operators,
the resolution of the stability issue must involve
nonrenormalizable operators that are irrelevant at low energies.
We return to this topic in section V.

\subsection{Microcausality}

A quantum field theory is microcausal
if any two local observables with spacelike separation commute.
In the Lorentz- and CPT-violating Dirac theory \rf{lagr},
the local quantum observables are fermion bilinears as usual,
and microcausality holds if
\beq
iS(x-x^{\prime})=\{\psi(x),
\overline{\psi}(x^{\prime})\}=0,
\quad (x-x^{\prime})^2<0.
\label{anticom}
\eeq
We work directly with the original field $\ps$ rather than $\ch$
because the observer Lorentz symmetry holds 
for the lagrangian \rf{lagr} written in terms of $\ps$,
whereas the conversion to $\ch$ is frame dependent.
Note that the anticommutator function $S(x-x^\prime)$
depends only on coordinate differences,
due to the translational invariance of the theory.

To investigate the conditions under which 
Eq.\ \rf{anticom} holds,
it is useful to obtain an integral representation for $S(x-x^\prime)$.
The latter can be found in terms of Green functions for the 
modified Dirac equation.
In the conventional case,
one usually starts with the Fourier decomposition of the field operators 
and proceeds by identifying spinor projection operators.
The latter are then expressed in terms of gamma matrices, 
the momentum, and the mass.
However,
in the present case
a straightforward generalization of this last step
is obstructed by the complexity of the modified Dirac equation.
Instead,
a more general argument can be adopted.

We proceed in a concordant frame.
First,
define the function
\beq
iG_R(x,x^{\prime})=
\Th(t-t^{\prime})\{\psi(x),
\overline{\psi}(x^{\prime})\},
\eeq
where $\Th$ denotes the usual Heaviside step function.
With the help of
the canonical anticommutators \rf{fieldcomm},
it can explicitly be checked that $G_R$ satisfies
\beq
(i\Ga^{\mu}{\prt}_{\mu}
-M)G_R(x,x^{\prime})=\delta^{(4)}
(x-x^{\prime}).
\eeq
It follows that $G_R(x,x^{\prime})$
is a Green function of the modified Dirac equation,
and therefore it can be written as
\beq
G_R(x,x^{\prime})=\int_{C_R}\fr{d^4\la}
{(2\pi)^4}\fr{e^{-i\la
\cdot(x-x^{\prime})}}
{\Ga^{\mu}\la_{\mu}-M} .
\label{rgreen}
\eeq
Inspection shows that $C_R$ is the contour
of the retarded Green function
passing above all poles in the complex $\la^0$ plane. 
Similarly,
it can be shown that the function defined by
\beq
iG_A(x,x^\prime)=
-\Theta(t^\prime-t)\{\psi(x),
\overline{\psi}(x^\prime)\}
\eeq
is the advanced Green function,
with the same representation as Eq.\ \rf{rgreen}
except that the contour $C_R$ is replaced with
a contour $C_A$ passing below all the poles.

The anticommutator function $S(x-x^\prime)$
can be written as $S=G_R-G_A$.
The integral represention for $S$
has the same form as Eq.\ \rf{rgreen}
except that $C_R$ is replaced by a contour $C$
encircling all poles in the clockwise direction.
If the matrix in the integrand of Eq.\ \rf{rgreen}
is explicitly inverted,
we can replace $\la^{\mu}\rightarrow i\prt^{\mu}$
in the matrix of cofactors 
cof$(\Ga^\mu\la_\mu-M)$
to obtain
\beq
S(z)=
{\rm cof}(\Ga^{\mu} i\prt_{\mu}-M)
\int_{C} \fr{d^4\la} {(2\pi)^4}
\fr{e^{-i\la\cdot z}} {\det(\Ga^{\mu}\la_{\mu}-M)} .
\label{ffgreen}
\eeq
The interchange of differentiation and integration
is justified because the contour can be deformed
so that the integrand is analytic
in the neighborhood of $C$ \cite{sg}.

Next, 
we take advantage of observer Lorentz invariance
and boost to a frame such that $z^{\mu}=(0,\vec{z})$.
The evaluation of $S(z)$ outside the light cone
is simplified when the spinor redefinition
discussed in Section IIIA 
can be performed in {\it all} observer frames.
A sufficient condition for this is:
\beq
c_{\mu \nu} = d_{\mu \nu} = e_\mu = f_\mu = g_{\la \mu \nu}=0 ,
\label{idnk}
\eeq
so that the derivative couplings take
the standard form with $\Ga^{\mu}=\ga^{\mu}$.
In this case, 
a hermitian hamiltonian always exists,
and the four poles of the integrand in Eq.\ \rf{ffgreen}
remain on the real axis in the complex $\la^0$ plane.

Under the condition \rf{idnk},
we can directly perform the contour integration
in Eq.\ \rf{ffgreen}.
For simplicity,
we assume here that all four roots
$E_{(j)}(\vec{p})$, $j=1,\ldots,4$,
of the dispersion relation are nondegenerate.
Cases with degenerate roots 
can be treated similarly with slight algebraic changes.
Explicit calculation yields
\begin{eqnarray}
&&\lefteqn{\int_C\fr{d\la^0}{2\pi}
\fr{1}{(\la^0-E_{(1)})(\la^0-E_{(2)})
(\la^0-E_{(3)})(\la^0-E_{(3)})}}
\nonumber\\
&&\qquad
\qquad
= \fr{i}{(E_{(1)}-E_{(2)})
(E_{(1)}-E_{(3)})(E_{(1)}-E_{(4)})}
\nonumber\\
&&\qquad
\qquad
\quad
+\fr{i}{(E_{(2)}-E_{(1)})(E_{(2)}
-E_{(3)})(E_{(2)}-E_{(4)})}
\nonumber\\
&&\qquad
\qquad
\quad
+\fr{i}{(E_{(3)}-E_{(1)})(E_{(3)}
-E_{(2)})(E_{(3)}-E_{(4)})}
\nonumber\\
&&\qquad
\qquad
\quad
+\fr{i} {(E_{(4)}-E_{(1)})(E_{(4)}
-E_{(2)})(E_{(4)}-E_{(3)})}
\nonumber\\
&&\qquad
\qquad
= 0 ,
\label{contour}
\end{eqnarray}
where the dependence of the $E_{(j)}$
on $\vec{p}$ has been suppressed.

This calculation shows that $S(z)$ vanishes outside the light cone
if (\ref{idnk}) is satisfied.
Thus, 
microscopic causality is ensured for the Dirac quantum field theory 
in the presence of Lorentz and CPT violation controlled by the coefficients
$a_{\mu}$, $b_{\mu}$ and $H_{\mu \nu}$.

The above argument can fail when Eq.\ \rf{idnk} is invalid.
For this more general case,
the poles of the integrand in Eq.\ \rf{ffgreen}
may no longer lie on the real $\la^0$ axis
in an arbitrary observer frame,
and the contour $C$ may therefore fail to encircle them all. 
This corresponds to the case where the bound on $\de^0$
discussed in section IIIA is violated,
so that the hamiltonian cannot be made hermitian 
and the roots of the dispersion relation can therefore be complex.

As an explicit example,
let us return to the $c_{00}$ model
with dispersion relation \rf{cdisp}
discussed in the previous subsection,
but without imposing $c_{00}>0$.
For this model,
the integration in (\ref{ffgreen})
can be performed analytically to yield:
\bea
S(z)&=&-(i\zeta\ga^0\prt^0
-i\ga^j\prt^j+m)
\nonumber\\
&&\qquad\qquad \times
\fr{1}{4\pi\zeta r}
\fr{\prt}{\prt r}[\Theta(w^2) J_0(m\sqrt{w^2})],
\label{prop}
\eea
where
$r=|\vec{z}|$,
$w^2=(z^0/\zeta)^2-\vec{z}^2$,
and $J_0(y)$ is the zeroth-order Bessel function.
Thus, 
the anticommutator function $S(z)$
vanishes only in the region defined by
$z^0<(1+c_{00})|\vec{z}|$.
Outside this region,
$S(z)$ could be nonzero.
Signal propagation therefore could occur with maximal speed
$1/(1+c_{00})$.
When $c_{00}$ is negative,
this exceeds 1 and hence violates microcausality.

To make further progress,
it is useful to 
introduce a definition of the velocity
of a particle valid for an arbitrary 3-momentum.
Even in the usual case without Lorentz and CPT violation,
the notion of a quantum velocity operator is nontrivial.
The presence of Lorentz and CPT violation further complicates the issue
\cite{ck}.
For definiteness,
we consider here the group velocity
defined for a monochromatic wave 
in terms of the dispersion relation
by Eq.\ \rf{vgr}.
This choice is appropriate for several reasons.
For one-particle states in the theory,
the flow velocities of the conserved momentum $P_{\mu}$ 
and the U(1) charge 
can be calculated from the corresponding conserved currents,
and they agree with the group velocity \rf{vgr}.
Also,
we have checked explicitly that 
$\langle d{\vec{x}}/dt\rangle =\vec{v}_{g}$
in the relativistic quantum mechanics of the $c_{00}$ model.
Moreover,
for the explicit examples considered above,
involving either no derivative couplings
or a $c_{00}$ coupling only,
the magnitude of the maximal attainable group velocity
is equal to the maximal speed of signal propagation
determined from the anticommutator function.

\begin{figure}
\centerline{\psfig{figure=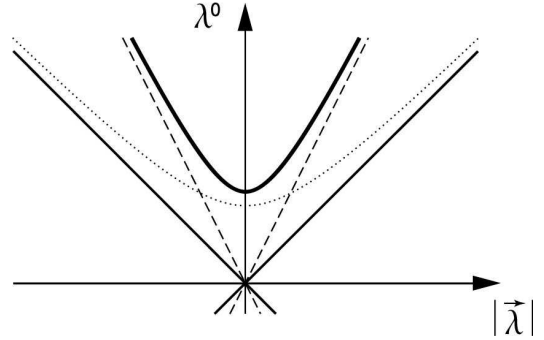,width=0.8\hsize}}
\smallskip
\caption{Dispersion relation for a model with only a 
large negative nonzero $c_{00}$ in a concordant frame.
The degenerate positive roots are displayed.
The dashed lines show their asymptotes.
The dotted line is the conventional dispersion 
for a massive particle.}
\label{fig3}
\end{figure}

\begin{figure}
\centerline{\psfig{figure=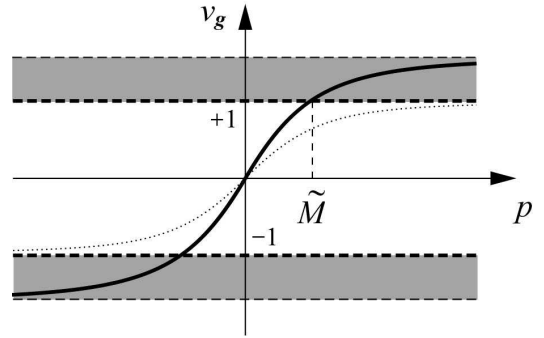,width=0.8\hsize}}
\smallskip
\caption{Group velocity for the dispersion relation 
of the model in Fig.\ 3 as a function of the 3-momentum
in a fixed direction.
The asymptotic development of velocities exceeding 1
is apparent in the shaded region,
which lies above a momentum scale $\tilde M$.
The heavy dashed lines correspond 
to the usual limiting velocities $\pm 1$. 
The dotted line is the usual result for a massive particle.}
\label{fig4}
\end{figure}

Figures 3 and 4 illustrate the situation for the $c_{00}$ model.
The dispersion relation in a concordant frame
is displayed in Fig.\ 3.
This figure shows that the maximal speed 
is attained asymptotically for large 3-momenta.
Figure 4 shows the group velocity 
as determined from the dispersion relation in the same frame.
Above a certain value $\tilde M$ of the 3-momentum magnitude,
all the group-velocity magnitudes exceed 1.

It follows from the above considerations that
a necessary condition to avoid microcausality violations 
is that the asymptotic behavior of the energy
must have a slope less than or equal to that of the usual light cone:
\beq
|\vec v_g| \leq 1,
\qquad 
|\vec{p}|\rightarrow\infty .
\label{cond2}
\eeq
Combined with Eq.\ \rf{cond1},
we see that a necessary condition for a positive root
to avoid both negative energies in some observer frame
and microcausality violations
is that the asymptotic behavior of the dispersion relation
must lie inside the forward light cone and satisfy
\beq
|\vec v_g| = 1,
\qquad 
|\vec{p}|\rightarrow\infty .
\label{cond3}
\eeq
Although this is only an asymptotic condition,
it nonetheless provides 
an interesting constraint on possible stable and causal models 
for Lorentz and CPT violation.

Insight about the scale $\tilde{M}$
of microcausality breakdown
can be obtained by determining the value of the 3-momentum 
at which the group velocity reaches 1:
$|\vec v_{g}| (|\vec p| =\tilde{M})=1$.
For the $c_{00}$ model,
the dispersion relation \rf{cdisp} gives 
\bea
\tilde{M}&=&\fr{\zeta}{\sqrt{1-\zeta^2}}m
\approx \fr{1}{\sqrt{-2c_{00}}}m+{\cO}(c_{00})
\nonumber\\
&\gsim &{\cO}(\sqrt{mM_P}) .
\label{mscale1}
\eea
In the last step,
the approximate equality holds
for a single suppression factor
$c_{00} \sim \cO(m/M_P)$.

The result \rf{mscale1} is a special feature of models with 
a nonzero $c_{\mu\nu}$ parameter.
It is the same as that for the case with $c_{00}>0$,
given in Eq.\ \rf{c00scale}.
We see that group velocities exceeding 1 occur 
in the $c_{00}$ model at energies well below
the scale $M_P$ of the underlying theory.
This may have physical implications,
as mentioned in the previous subsection.

To see what happens for other Lorentz- and CPT-violating terms 
with derivative couplings,
consider a model with only a nonzero $e_{\mu}$ term.
Its dispersion relation is 
\beq
\la^2-(m-\la\cdot e)^2=0 .
\label{edisp}
\eeq
For simplicity,
we take $e_{\mu}$ to be timelike
and choose the concordant frame to have $\vec{e}=0$.
The scale $\tilde M$ of microcausality violation
is then found to be 
\bea
\tilde{M}&=&\fr{1}{e_0}m
\nonumber\\
&\gsim &{\cO}(M_P) ,
\label{mscale2}
\eea
where in the last step 
the approximate equality is attained for
a single Planck-scale suppression factor,
$e_{0} \sim \cO(m/M_P)$,
as before.
This confirms that microcausality is violated
in the $e_{\mu}$ model at the scale of the underlying theory,
as expected.

The $e_\mu$ model can also be used 
to illustrate the relation between microcausality
and hermiticity of the hamiltonian $H$.
In the $e_{\mu}$ model,
the matrix $\ga^0\Ga^0$ takes the explicit form
\beq
\ga^0\Ga^0=\left(\begin{array}{cccc}
1+e_0 & 0 & 0 & 0\\
0 & 1+e_0 & 0 & 0\\
0 & 0 & 1-e_0 & 0\\
0 & 0 & 0 & 1-e_0
\end{array}
\right)
\label{explgam}
\eeq
in the Pauli-Dirac representation.
Provided $|e_0|<1$, 
the spectrum of $\ga^0\Ga^0$ containes
positive numbers only,
a matrix $A$ satisfying Eq.\ \rf{redef} can be found,
and a hermitian hamiltonian $H$ exists.
However, 
if $|e_0|>1$,
two eigenvalues become negative,
$\ga^0\Ga^0$ is no longer congruent to the identity,
the spinor-redefinition matrix $A$ cannot exist,
and a hermitian $H$ cannot be found.

The same problem is reflected at the level
of the dispersion relation \rf{edisp}.
Its solutions
\beq
\la^0_\pm=\fr{e_0(m+\vec{\la}
\cdot\vec{e})\pm
\sqrt{(m+\vec{\la}\cdot\vec{e})^2
+(1-{e_0}^2)\vec{\la}^2}}{{e_0}^2-1}
\label{roote}
\eeq
can become complex for $|e_0|>1$.
Since it is always possible to find an observer frame
in which this condition is satisfied,
the model is inconsistent with observer invariance
of the hermiticity of $H$.
This again indicates that the argument
for microcausality can fail when the condition \rf{idnk} is invalid.

\begin{figure}
\centerline{\psfig{figure=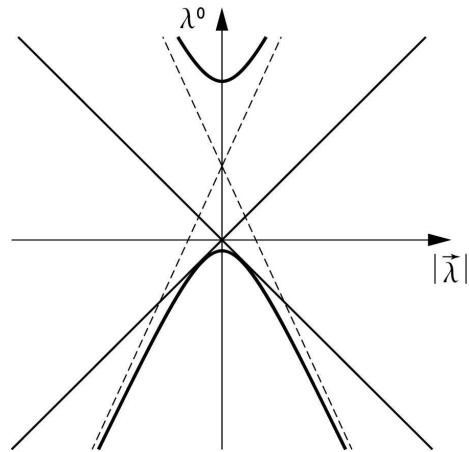,width=0.7\hsize}}
\smallskip
\caption{Dispersion relation for a model with only a 
large nonzero $e_{0}$ in a concordant frame.
One positive root and its negative partner are displayed.
The dashed lines show the asymptotes.}
\label{fig5}
\end{figure}

\begin{figure}
\centerline{\psfig{figure=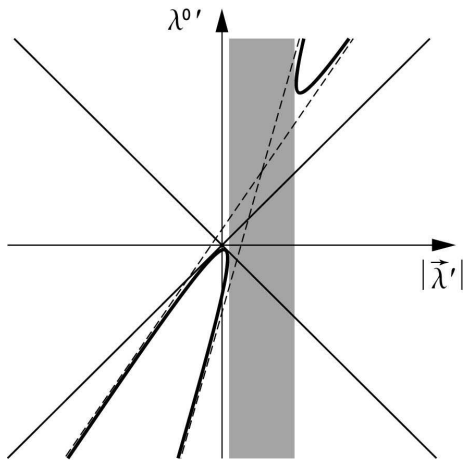,width=0.7\hsize}}
\smallskip
\caption{Dispersion relation for the model of Fig.\ 5
as seen by an observer strongly boosted 
relative to the concordant frame.
The occurrence of multiple-valued energies for a given root is apparent.
The positive root and its negative partner have no
real values of the energy for 3-momenta in the shaded region.
The dashed lines show the asymptotes.}
\label{fig6}
\end{figure}

Figures 5 and 6 illustrate
in the context of the $e_\mu$ model
how eigenenergies can be real in one observer frame
and complex in another,
despite the observer invariance of the dispersion relation.
Figure 5 shows the dispersion relation for a model 
with a nonzero $e_0$ only,
in a concordant frame.
One of the two positive roots and its negative partner are displayed.
The eigenenergies are real for all 3-momenta.
However,
the slope of the dispersion relation exceeds 1 for
a sufficiently large 3-momentum.
The effect of this on a positive root and its negative partner 
as seen by an observer in a strongly boosted frame 
is displayed in Figure 6.
These two roots admit no real value of the energy 
for 3-momenta in the shaded region.
Moreover,
there is a range of 3-momenta for which the
dispersion relation has multiple-valued roots.

This feature can be expected in the general case,
whenever the magnitude $|\vec v_g|$ of 
the slope of the dispersion relation
in a concordant frame exceeds 1.
More generally,
the individual branches of the dispersion relation 
should remain one-to-one mappings under observer Lorentz transformations,
so that each 3-momentum has exactly one image point.
The number of real solutions to the dispersion relation 
is then invariant under observer boosts.
In terms of the asymptotic behavior of the dispersion relation
in the general case,
we see that the existence requirements 
for the spinor redefinition \rf{redef}
and for a hermitian hamiltonian $H$
also lead to the condition \rf{cond2}. 

The above analysis reveals that difficulties
with causality in the Lorentz- and CPT-violating Dirac theory 
arise primarily for Planck-scale 4-momenta in a concordant frame
or for observers undergoing a Planck boost relative to this frame.
Nonetheless,
it would be theoretically interesting 
to have a framework for Lorentz and CPT violation
in which microcausality is exactly preserved.
Moreover,
constraints from the requirement of causality
may offer insight into the nature of an underlying theory
with Lorentz and CPT violation.
This is the subject of the following section.

\section{Planck-Scale Effects}

The results of the previous section indicate that 
a quantum field theory of massive fermions with terms 
containing explicit Lorentz and CPT violation 
generically develops difficulties with stability or causality. 
However,
if the coefficients controlling the violation are Planck-suppressed,
as in the standard-model extension,
the difficulties arise only at high energies or high boosts 
determined by the Planck scale.

Many possible sets of values of the coefficients 
$a_{\mu}$, $b_\mu$, $\ldots$, $H_{\mu\nu}$
for Lorentz and CPT violation
in Eq.\ \rf{lagr} eliminate one of the two difficulties.
However,
we are unaware of any combination of the coefficients
that simultaneously maintains both stability and causality.
Although it is conceivable that a satisfactory combination
would be naturally selected by a mechanism for Lorentz and CPT breaking
in an underlying theory,
we conjecture that no such combination exists.
A definitive argument to settle this issue
would be of interest
but appears hampered by the complexity of the 
dispersion relation \rf{convenientform}.

We have previously advocated spontaneous Lorentz and CPT breaking
in a Lorentz-covariant theory at the Planck scale
as a possible mechanism that could generate 
the apparent Lorentz and CPT violations at low energies
\cite{str,kp}.
Indeed,
the standard-model extension includes by construction
all possible renormalizable terms maintaining the usual gauge structure 
while potentially originating in spontaneous Lorentz breaking. 
This reasoning is a top-down approach,
with theoretical considerations at the Planck scale 
suggesting that spontaneous Lorentz violation might emerge 
as the apparent violation in the standard-model extension. 
However,
the requirements of stability and causality 
appear strong enough to adopt the inverse line of reasoning.
Thus,
as the Planck scale is approached,
higher-order nonrenormalizable operators 
coming from the fundamental theory should play an increasing role.
The structure of the standard-model extension 
as a conventional quantum field theory
should therefore undergo a corresponding modification,
which could provide insight into the nature of the 
fundamental theory at the Planck scale.
In the remainder of the present section,
we fill in some details for this set of ideas.

\subsection{Spontaneous Lorentz and CPT breaking}

Since a theory with spontaneous Lorentz and CPT violation
starts from a Lorentz-invariant lagrangian 
and hence has Lorentz-covariant dynamics,
it is unsurprising that it avoids at least some of  
the difficulties plaguing more general models 
involving Lorentz and CPT violation.
For example,
one consequence of spontaneous violation
is the natural maintenance of observer Lorentz invariance,
which the previous sections have shown to be an important advantage.
Thus,
given a lagrangian invariant under both
observer and particle Lorentz transformations,
spontaneous symmetry breaking violates only the latter.
The point is that observer Lorentz invariance is a statement about
physical behavior under certain coordinate changes 
made by an independent external observer,
and once this property is built into a theory
it cannot be removed by the behavior of fields internal to the theory.
In contrast,
imposing observer Lorentz invariance
in a theory with explicit Lorentz breaking
requires an additional \it ad hoc \rm choice.

Spontaneous violation manifests itself physically
because the Fock-space states 
are constructed on a noninvariant vacuum.
Any difficulties with spontaneous Lorentz and CPT violation
must therefore be a consequence of 
Lorentz- and CPT-violating properties of the ground state.
However,
the link between stability, causality, and Lorentz symmetry 
does indeed depend in part on the notion of an invariant vacuum.
The difficulties uncovered in the previous section
can be regarded as a consequence of vacuum noninvariance.
For example,
the vacuum state in one frame is not necessarily
the lowest-energy state in all frames.
Despite its advantages,
one therefore might expect that 
spontaneous Lorentz and CPT violation alone 
may be insufficient to guarantee stability and causality
at all scales in a generic quantum field theory.

To gain insight into this issue,
it is useful to consider a toy quantum field theory
describing a Dirac fermion $\ps$ 
interacting with a vector field $B_\mu$,
with a potential for the vector 
that induces spontaneous Lorentz and CPT violation
\cite{ks}.
The lagrangian is 
\bea
{\cal L}&=&\overline{\psi}(\frac{1}{2}i
\ga^{\mu}
\stackrel{\leftrightarrow}{\prt}_{\mu}
-m-\xi \ga_5 \ga^{\mu} B_\mu)\psi
\nonumber\\
&& \qquad 
-\frac 1 4 F^{\mu\nu}F_{\mu\nu}
-\frac 1 4 \la (B^\mu B_\mu -\be^2)^2.
\label{toylagr}
\eea
The fermion $\ps$ has mass $m$ and is chirally coupled
to the vector $B_\mu$ with dimensionless strength $\xi$.
The field strength $F_{\mu\nu}$ for $B_\mu$ 
is defined as $F_{\mu\nu} = \prt_\mu B_\nu - \prt_\nu B_\mu$,
as usual,
while the potential term for $B_\mu$ is controlled
by a dimensionless constant $\la$
and by a constant $\be$ with dimensions of mass
satisfying $\be^2 > 0$.

The lagrangian \rf{toylagr}
is a scalar under both observer and particle Lorentz transformations 
and contains no explicit Lorentz- and CPT-violating terms.
However,
the last term triggers a Lorentz- and CPT-violating 
vacuum expectation value
$\vev{B_\mu} = \be_\mu$,
where $\be_\mu$ is a constant 4-vector
satisfying $\be_\mu \be^\mu = \be^2$.
Note the close analogy to spontaneous symmetry breaking
in the standard O($N$) model with $N=4$.
The Lorentz invariance of the lagrangian \rf{toylagr}
means that the constant vector $\be_\mu$ can be arbitrarily chosen,
but a definite choice must be specified to establish the quantum physics. 
This choice forces the particle Lorentz symmetry 
to be spontaneously broken on the Fock space.

The physics of interest is described by fluctuations
about the vacuum. 
Redefining $B_\mu\to \be_\mu+B_\mu$ 
in parallel with the usual case yields
\bea
{\cl}&=&
\overline{\psi}
[\frac{1}{2}i
\ga^{\mu}
\stackrel{\leftrightarrow}{\prt}_{\mu}
-m-\xi \ga_5 \ga^{\mu} (\be_\mu+B_\mu)]
\psi
\nonumber\\
&& 
-\frac 1 4 F^{\mu\nu}F_{\mu\nu}
-\frac 1 4 \la (B^\mu B_\mu -2B\cdot \be)^2
\nonumber\\
&=&
\overline{\psi}
(\frac{1}{2}i 
\ga^{\mu}
\stackrel{\leftrightarrow}{\prt}_{\mu}
-m-\ga_5 \ga^\mu b_\mu) \psi
+ {\cl}^\prime ,
\label{shiftlagr}
\eea
where in the last step we have identified
$\xi \be_\mu$ with $b_\mu$
and explicitly displayed all the quadratic fermion terms in $\cl$. 
The remaining piece ${\cl}^\prime$ of the lagrangian
contains only bosonic quadratic terms and interactions.
We see that the spontaneous Lorentz and CPT violation 
in the lagrangian \rf{toylagr}
has generated the $b_\mu$ model discussed in previous sections.

The free-field Fock space of the quantum theory 
associated with $\cl$
contains one-fermion states determined by 
the quadratic terms in Eq.\ \rf{shiftlagr}.
These states have dispersion relations 
given by Eq.\ \rf{bdisp},
as before.
They therefore suffer from the same problems 
of instability as the $b_\mu$ model discussed in section IVB.
This leads to difficulties within
the standard framework of perturbative quantum field theory,
since the interacting fields are normally constructed
iteratively from the free fields
under the assumption that the effects of interactions are small. 
The toy model therefore still has interpretational difficulties,
despite the spontaneous nature of the Lorentz and CPT violation.

A similar argument applies to more general models.
Since the theory described by Eq.\ \rf{lagr}
contains the most general terms quadratic in the fermion fields 
and arising in a renormalizable theory,
any conventional fermion field theory 
with spontaneous Lorentz and CPT violation
analogous to Eq.\ \rf{toylagr}
must generate free-fermion Fock-space states
with dispersion relations contained as a subset
of Eq.\ \rf{convenientform}.
If all such dispersion relations indeed lead to
either stability or causality violations at some large scale,
as expected from the discussion in section IV,
then it follows that no conventional lagrangian of fermions 
with spontaneous Lorentz and CPT violation
has a completely satisfactory perturbative quantum field theory. 
Although it is conceivable that a nonperturbative analysis
taking the full structure of the theory into account
would reveal a consistent theory 
satisfying stability and causality,
this appears unlikely.
Even this possibility is excluded 
if the quantum field theory is \it defined \rm
in terms of its perturbative expansion,
as is sometimes done in the literature.

The above discussion shows that
spontaneous symmetry breaking in a conventional quantum field theory
can naturally generate Lorentz- and CPT-violating terms
of the form in \rf{lagr}
and ensures various desirable features
such as observer Lorentz symmetry.
Provided the coefficients for Lorentz and CPT violation are small,
as in the standard-model extension,
difficulties arise only at large scales.
However,
by itself spontaneous Lorentz violation
is insufficient to ensure stability and causality 
at energies determined by the Planck scale.
Maintaining stability and causality requires an additional ingredient
that goes beyond conventional quantum field theory.
This is consistent with the idea that
the observation of Lorentz and CPT violation 
would provide a unique signal of Planck-scale physics.

\subsection{Nonlocality}

If indeed the requirements of stability and causality
are to be satisfied by free-field terms,
then it is of interest to identify a class of theories for which
no difficulties arise in the quadratic lagrangian.
Such theories would need to include terms beyond 
the ones in Eq.\ \rf{lagr}.
The new terms must be nonrenormalizable,
and in a realistic scenario with spontaneous Lorentz violation 
they would correspond to higher-order nonrenormalizable operators 
correcting the standard-model extension at energies
determined by the Planck scale.

The first step is to determine whether any type of dispersion relation
can satisfy all the requirements for consistency.
In a concordant frame,
a satisfactory dispersion relation describing Lorentz and CPT violation
would reproduce the physics of Eq.\ \rf{convenientform}
for small 3-momenta
but would avoid spacelike 4-momenta and group velocities exceeding 1
for large 3-momenta.
Moreover,
its asymptotic behavior would need to obey Eq.\ \rf{cond3}.
These requirements could be implemented 
by combining the coefficients for Lorentz and CPT violation
with a suitable factor suppressing them only at large 3-momenta.
A factor of this type must be essentially constant at small 3-momenta
and must overwhelm polynomial powers at large 3-momenta. 
Since the distinction between small and large 3-momenta
is a frame-dependent concept,
it is to be expected that a suitable factor
would also be frame-dependent
and hence involve Lorentz- and CPT-violating coefficients.

A complete treatment of the possibilities lies outside the 
scope of the present work.
Instead,
we prove by example that suitable dispersion relations 
can in principle exist 
by providing explicit situations with the desired features.
We present here two cases 
that are closely related to the $b_\mu$ and $c_{\mu\nu}$
models discussed in section IV.
To simplify the discussion,
we disregard here issues associated with 
the size of the coefficients for Lorentz and CPT violation
and take all masses and Lorentz- and CPT-breaking coefficients 
to be of order 1
in appropriate units. 
This permits a focus on resolving
the problems of stability and causality
at Planck-scale energies in a concordant frame
without the complications introduced by the hierarchy of scales.

Consider first a dispersion relation obtained from
Eq.\ \rf{bdisp} for the $b_{\mu}$ model 
by combining all appearances of $b_\mu$ with
an appropriate exponential factor.
For simplicity,
we take a model with only a nonzero $b_0$ 
in a concordant frame.
Multiplication of each factor of $b_0$ by
$\exp[-(b_0 \la_0)^2]$
suppresses the effect of $b_0$ at high energies
with minimal effect at low energies.
In an arbitrary frame,
observer Lorentz invariance implies the resulting 
modified dispersion relation takes the form 
\bea
&&[\la^2-b^2\exp[-2(b\cdot \la)^2] -m^2]^2
\nonumber\\
&& \quad 
+4b^2\la^2\exp[-2(b\cdot \la)^2] 
\nonumber\\
&& \quad 
-4(b\cdot\la)^2\exp[-2(b\cdot \la)^2]=0.
\label{bdispmod}
\eea
For $b_0$ of appropriate size,
the positive roots of this modified dispersion relation 
remain positive in all frames.
This provides a proof by example
that a suitable modification of the dispersion relation
can be found that removes the 
difficulty with stability in arbitrary frames.

Figure 7 shows the dispersion relation for the 
modified $b_\mu$ model
in the special case where only $b_0$ is nonzero
in a concordant frame.
At small energies,
the exponential factors are negligible 
and the behavior is essentially like
that of the original $b_0$ model.
However,
at large energy the exponential factors dominate,
causing the dispersion relation to remain within
the light cone while asymptotically approaching it
as required by condition \rf{cond3}.
The modified $b_\mu$ dispersion relation \rf{bdispmod} 
therefore has no difficulties with energy positivity in any frame.

To establish that microcausality is also preserved,
the group velocity of the modified dispersion relation \rf{bdispmod} 
can be examined.
Figure 8 shows that the group velocity can indeed lie 
between the usual limiting values $\pm 1$
for all values of the 3-momentum 
despite the modification to the dispersion relation. 
Note that the asymmetry of this plot reflects the asymmetry of
the corresponding curve in Fig.\ 7.

It is also possible to find examples where the difficulties
with causality are absent.
For example,
consider the dispersion relation obtained from
Eq.\ \rf{cdisp} for the $c_{00}$ model with $c_{00}<0$ 
by multiplying each factor of $c_{00}$
with an exponential factor $\exp(c_{00} \la_0^2)$.
In an arbitrary frame,
the result is a modification of Eq.\ \rf{cdispcov} given by 
\bea
&&(\et_{\al\mu}+c_{\al\mu} \exp(c_{\be\ga}\la^\be\la^\ga))
\nonumber\\
&&\qquad\times
(\et^\al_{\pt{\nu}\nu}+c^\al_{\pt{\nu}\nu} \exp(c_{\be\ga}\la^\be\la^\ga))
\la^\mu\la^\nu
-m^2 = 0.
\label{cdispmod}
\eea
The exponential factors remove the microcausality violations
that previously occurred at large $\la_\mu$.
Indeed,
it can be shown that the group velocity remains below 1
for all values of $\vec \la$.
This proves by example
that a suitable modification of the dispersion relation
can eliminate difficulties with microcausality
\cite{fn4}.

\begin{figure}
\centerline{\psfig{figure=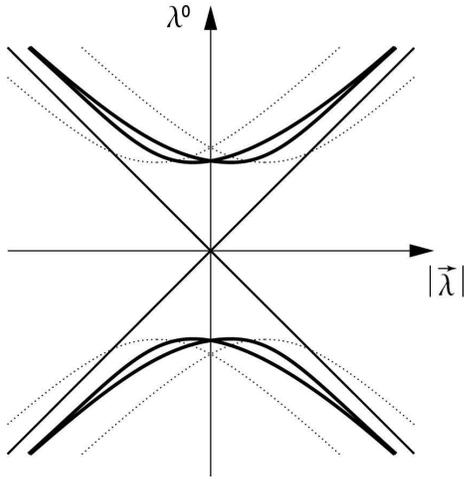,width=0.7\hsize}}
\smallskip
\caption{Dispersion relation for a model with 
only a large nonzero $b_0$ in a concordant frame
and exponential suppression at large energy.
All four roots are displayed.
None cross the light cone.
The dotted lines are the four roots for the
$b_0$ model \it without \rm the exponential suppression.}
\label{fig7}
\end{figure}

\begin{figure}
\centerline{\psfig{figure=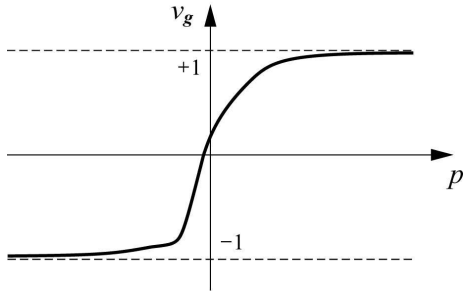,width=0.7\hsize}}
\smallskip
\smallskip
\caption{Group velocity for the dispersion relation 
of the model in Fig.\ 7 as a function of the 3-momentum 
in a fixed direction.
The modified dispersion has no group velocity exceeding 1. 
The dashed lines correspond to the usual limiting velocities $\pm 1$.}
\label{fig8}
\end{figure}

Figure 9 displays the dispersion relation for the special
case of a modified model with only a nonzero $c_{00}$
in a concordant frame.
At small energies,
the exponential factors are negligible 
and the behavior is essentially like  
that of the original $c_{00}$ model.
However,
at large energy the exponential factors dominate,
so the group velocities never exceed 1
and causality is maintained.
The asymptotes of the dispersion relation coincide with the light cone,
as required by Eq.\ \rf{cond3}.
The group velocity of the modified dispersion relation \rf{cdispmod} 
is shown as a function of the 3-momentum in Fig.\ 10. 
It remains within the usual limiting velocities everywhere, 
as desired.

\begin{figure}
\centerline{\psfig{figure=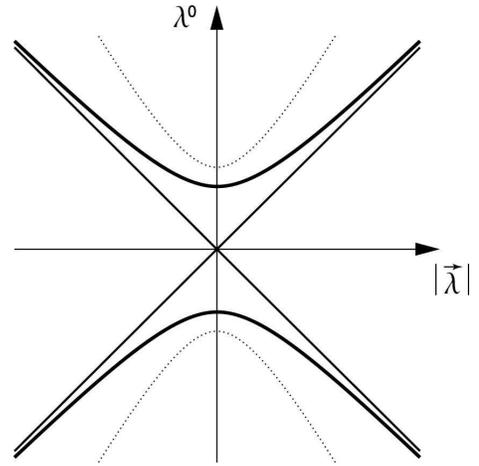,width=0.7\hsize}}
\smallskip
\caption{Dispersion relation for a model with only a 
large nonzero $c_{00}$ in a concordant frame
and exponential suppression at large energy.
Only two curves appear because there is 
a two-fold degeneracy among the four roots.
The dotted lines are the corresponding roots for the
$c_{00}$ model \it without \rm the exponential suppression.}
\label{fig9}
\end{figure}

\begin{figure}
\centerline{\psfig{figure=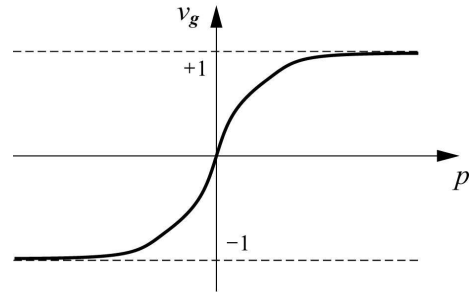,width=0.7\hsize}}
\smallskip
\caption{Group velocity for the dispersion relation 
of the model in Fig.\ 9 as a function of the 3-momentum 
in a fixed direction.
The modified dispersion has no group velocity exceeding 1. 
The dashed lines correspond to the usual limiting velocities $\pm 1$.}
\label{fig10}
\end{figure}

The above demonstrations prove that dispersion relations 
violating Lorentz and CPT while maintaining stability and causality
can exist.
It would be of interest to identify theories 
from which these dispersion relations emerge naturally.
The appearance of transcendental functions of the momenta
corresponds to the occurrence of 
derivative couplings of arbitrary order in the lagrangian.
A satisfactory theory with Lorentz and CPT violation
appears necessarily to be nonlocal in this sense.
Although it is conceivable that 
a theory with explicit Lorentz breaking might satisfy the requirements
of stability and causality,
it would appear somewhat contrived to
implement both the necessary observer Lorentz invariance
and nonlocal couplings by hand.
In contrast,
we see that spontaneous Lorentz and CPT violation in a nonlocal theory 
can naturally yield the desired ingredients
for stability and causality at all scales.

\subsection{String theory}

Our field-theoretic considerations 
seeking the nature of Planck-scale corrections
to a low-energy quantum field theory with Lorentz and CPT violation 
have thus led naturally to the case 
of a nonlocal theory with spontaneous symmetry breaking.
String theories have nonlocal interactions,
and it is of interest to determine whether they
could be of the desired kind.
Although a satisfactory realistic string theory
has yet to be formulated,
string field theories do exist for some simple string models
and have already been used to investigate 
microcausality in the Lorentz-invariant case
\cite{strcaus}.
Moreover,
studies of string field theory provided the original motivation
for identifying spontaneous Lorentz and CPT violation as a serious 
candidate signal from the Planck scale
\cite{str}
and for the construction of the standard-model extension
as the appropriate low-energy limit.

In the remainder of this section,
we examine the structure of 
the field theory for the open bosonic string
to see whether it is compatible 
with dispersion relations of the desired type.
Although this theory is unrealistic in detail,
the structural features of interest
are generic to string field theories
and so provide insight into the possibility of
generating a consistent theory with
spontaneous Lorentz and CPT violation.

The open bosonic string has no fermion modes,
so instead we focus on the dispersion relation for the scalar tachyon mode
in the presence of Lorentz- and CPT-violating expectation values 
of tensor fields.
In general,
the analogue of Eq.\ \rf{lagr} 
for a single real massive scalar field $\ph$ is
\cite{ck}
\beq
\cl = 
\half \prt_\mu \ph \prt^\mu\ph
- \half m^2 \ph^2
+ \half k_{\mu\nu} \prt^\mu \ph \prt^\nu\ph.
\label{scalarlagr}
\eeq
Here, 
$k_{\mu\nu}$ is a dimensionless 
coefficient for Lorentz violation that preserves CPT.
It can be taken as real, symmetric, and traceless.
The dispersion relation for this theory 
is closely related to that for the lagrangian \rf{lagr}
with a nonzero coefficient $c_{\mu\nu}$ only.
For the special case with only $k_{00}$ nonzero 
in a concordant frame,
the dispersion relation of the theory \rf{scalarlagr}
is just that in Eq.\ \rf{cdisp}
with the identification $\ze^2=1+k_{00}$.
Studying the dispersion relation of the scalar tachyon mode
in the presence of Lorentz violation 
is therefore more appropriate 
than might perhaps be expected \it a priori. \rm

The action for the Witten string field theory
\cite{ew}
can be written in the Chern-Simons form 
\beq
I(\Ps) =
\fr 1 {2\ap} \int \Ps \star Q\Ps
+ \fr g 3 \int \Ps \star \Ps \star \Ps,
\label{action}
\eeq
where $\al^\prime$ is the Regge slope 
and $g$ is the on-shell 3-tachyon coupling
at zero momentum.
The operator $Q$ acts as a quadratic kinetic operator.
The interactions are controlled by the star operator $\star$,
which joins the left half of one string to the right half of another.
The integral joins the left half of a string onto its own right half.

The vibrational modes of the string are the particle states.
The field $\Ps$ can be decomposed as a linear combination
of ordinary particle fields
with coefficients that are solutions of the first-quantized theory,
expressed as creation operators $\al_{-1}$, $\ldots$
acting on a vacuum $\ket{0}$.
Following the notation of
Ref.\ \cite{ksobscs},
the fields in $\Ps$ are found to include among others 
a scalar $\ph$ (the tachyon)
and a series of $2j$-tensors $B_{\mu\nu}$, $D_{\mu\nu\rh\si}$, $\ldots$:
\bea
\Ps &=& 
\left(\ph + \ldots 
+ \fr 1 {\sqrt{2}} B_{\mu\nu}\al_{-1}^\mu\al_{-1}^\nu
\right.
\nonumber\\
&& \qquad
\left.
+ \fr 1 {2\sqrt{6}} D_{\mu\nu\rh\si}
\al_{-1}^\mu\al_{-1}^\nu\al_{-1}^\rh\al_{-1}^\si
+ \ldots\right) 
\ket{0}.
\eea

The explicit lagrangian for the theory in terms of particle fields
to low orders has been obtained in
Ref.\ \cite{ksobscs}.
Our interest here lies merely in 
determining whether the theory can in principle
contain the types of term
necessary for a stable and causal dispersion relation
involving Lorentz violation.
We therefore proceed under the assumption 
that spontaneous Lorentz violation has occurred,
possibly along the lines discussed in Ref.\ \cite{str},
and has generated nonzero expectation values for the $2j$-tensors:
$\vev{B_{\mu\nu}}$, $\vev{D_{\mu\nu\rh\si}}$, $\ldots$.
Note that this assumption preserves CPT,
as desired.

Follow the approach of subsection VA,
we directly extract relevant quadratic terms in the lagrangian 
involving the tachyon.
This procedure yields the lagrangian 
\bea
\cl &\supset &
\half \prt_\mu \ph \prt^\mu \ph  
+ (\al^{\prime -1} + k_0) \ph^2
+ \ldots
\nonumber\\
&&\qquad
+ k_1 \vev{B_{\mu\nu}} \prt^\mu \ph \prt^\nu \ph  
+ \ldots
\nonumber\\
&&\qquad
+ k_2 \vev{D_{\mu\nu\rh\si}} 
\prt^\mu \ph \prt^\nu \ph \prt^\rh \ph \prt^\si \ph  
+ \ldots .
\eea
Here,
the scalar parameters $k_0$, $k_1$, $k_2$, $\ldots$
are fixed by the theory,
but their specific values are irrelevant 
for the present considerations.
Each ellipsis represents quadratic terms involving
other tensor expectation values 
and terms with powers of $\la^2$.

For a plane-wave tachyon solution,
the dispersion relation resulting from this lagrangian
takes the form
\bea
&&\la^2 + (\al^{\prime -1} + k_0)
+ \ldots 
+ k_1 \vev{B_{\mu\nu}} \la^\mu \la^\nu   
+ \ldots 
\nonumber\\
&&\qquad
+ k_2 \vev{D_{\mu\nu\rh\si}} 
\la^\mu \la^\nu \la^\rh \la^\si 
+ \ldots = 0.
\label{tachdisp}
\eea
We see that the structure of this equation does indeed contain features
similar to those needed for a dispersion relation
satisfying criteria for stability and causality.
Thus,
for example,
the type of term in the toy dispersion relation \rf{cdispmod}
is a subset of the terms displayed in Eq.\ \rf{tachdisp},
when only 0th components of the $2j$ tensors are nonzero 
and the $2j$th-tensor expectation value is proportional
to $(k_{00})^{j}$.

We emphasize that the purpose of the above discussion
is only to provide an outline indicating how
an acceptable dispersion relation for Lorentz violation
might emerge in the context of string theory.
In particular,
we make no claim that the tachyon itself
must \it necessarily \rm obey such a relation,
although it is conceivable that it does
\cite{str}.
Here,
the tachyon dispersion relation is used 
merely as an example to display explicitly
the appearance of nonlocal couplings in string theory 
that could be appropriate for 
a stable and causal theory with spontaneous Lorentz violation.
Such couplings are generic both for other fields in
the open bosonic string 
and for fields in other string theories,
including ones with fermions.

It would be of interest to find an explicit 
analytical construction for a Lorentz-violating solution
in some string field theory
and demonstrate its stability and causality.
The most accessible case is likely to be
the open bosonic string,
but other string field theories
with fermions could be amenable to investigation.
If such a solution exists,
it may be possible to find it using  
the methods of Ref.\ \cite{kp00}.
These interesting issues lie beyond the scope of the present work.

\section{Summary}

In this paper,
we have investigated the issues of stability and causality
in quantum field theories incorporating Lorentz and CPT violation.
No difficulties arise at low energies
provided the coefficients for Lorentz violation are small.
However,
local quantum field theories of fermions 
involving Lorentz violation
generically develop difficulties with either stability or causality
at some scale in every inertial frame.

On experimental and theoretical grounds,
it is to be expected that 
the parameters controlling the Lorentz and CPT violation are
Planck suppressed in any Earth-based laboratory frame.
In this physical situation,
except for a special case involving a scale intermediate 
between the low-energy and the Planck scales,
the difficulties appear only for particles with Planck-scale energies 
or in inertial frames undergoing Planck-scale boosts.
In particular,
the detailed analysis can be applied to the 
fermion sector of the standard-model extension,
which is thereby seen to have a regime of validity
comparable in many respects 
to that expected for the usual standard model.
The high-energy difficulties 
are characterized by one-particle dispersion relations
with tails either crossing the light cone
or developing group velocities exceeding 1.
The former result in instabilities,
while the latter produce microcausality violations.

As part of the analysis,
we have presented the relativistic quantum mechanics
and the quantum field theory of a massive fermion 
governed by the quadratic sector of a renormalizable lagrangian
with general Lorentz- and CPT-violating terms.
Much of the discussion can be extended to  
quadratic terms in a quantum field theory for a massive scalar
with Lorentz and CPT violation,
by virtue of the generality 
of the dispersion relation \rf{convenientform}
and the usual type of connection 
between the Dirac and Klein-Gordon equations.
Some of the results should also apply to the case of 
massless particles,
including any massless neutrinos and the photon or other gauge bosons.
However,
further effort is likely to be required to account correctly
for the differences between massive and massless representations
of the Lorentz group and for the effects of gauge symmetry.
Our methodology and general results are also applicable 
to nonrenormalizable terms in an effective theory.
The limitation to renormalizable terms
in our analysis is largely a matter of convenience,
chosen to minimize complications
in the identification of the origin and resolution 
of the difficulties with Lorentz and CPT violation.

The issues with stability and causality 
can be resolved under suitable circumstances.
An important ingredient in this 
is the requirement of observer Lorentz invariance,
which is guaranteed if the Lorentz and CPT violation 
develops spontaneously 
in a Lorentz-covariant underlying theory.
This provides a link between the Fock spaces constructed
by different inertial observers.
In contrast,
in theories based on explicit Lorentz violation instead,
this condition must either be imposed by hand 
or be replaced by some other \it ad hoc \rm condition.

We have shown explicitly that spontaneous Lorentz and CPT violation 
in suitable nonlocal theories can generate dispersion relations
avoiding the problems with stability and causality.
In particular,
the necessary structures 
appear in the context of string field theories.
We find it noteworthy that imposing stability and causality 
on quantum field theories with Lorentz violation
leads naturally both to insight about the nonrenormalizable terms
emerging as the Planck scale is approached 
and to requirements compatible with string field theories.
This reverses the usual chain of reasoning
by which spontaneous Lorentz and CPT violation
in some fundamental theory leads to 
the standard-model extension in the low-energy limit
where nonrenormalizable terms become irrelevant. 

The analysis in this work supports the idea 
that a stable and causal realistic fundamental theory
involving spontaneous Lorentz and CPT violation exists.
If so,
it would lead to potentially observable effects at sub-Planck energies
described by the Lorentz- and CPT-violating standard-model extension.
This offers the promising possibility
of providing a unique experimental signature
of Planck-scale physics. 

\section*{Acknowledgments}
This work is supported in part 
by the United States Department of Energy
under grant number DE-FG02-91ER40661.

\begin{appendix}

\section{Bound for $\de^0$}

The key to bounding $\de^0$ is to obtain a bound on
$\det (\ga^0\Ga^0)
=\det (I+\ep^0)$
in terms of the components of the matrix 
$\ep^0$ controlling the Lorentz and CPT violation.
Expanding the determinant yields 4!=24 terms,
each a product of 4 matrix elements of $I+\ep^0$.
It can be written 
\beq
\det (I+\ep^0)=
(1+\ep^0_{11})(1+\ep^0_{22})
(1+\ep^0_{33})(1+\ep^0_{44}) + \ldots,
\label{det}
\eeq
where $\ep^0_{jk}$ denotes the $jk$ element of $\ep^0$
and the ellipsis represents the 23 remaining terms,
none of which are at zeroth order in $\ep^0$.

Define
$\ep=\max_{j,k}\{|\ep^0_{jk}|\}$,
the matrix element
with the largest absolute value.
Then, 
a lower bound for the term displayed in the expansion \rf{det}
is $(1-\ep)^4$.
Provided $\ep<\frac{1}{2}$,
the largest of the remaining terms
is bounded above by $\ep(1+\ep)^3$.
It follows that
\beq
\det (I+\ep^0) \ge (1-\ep)^4 -23\ep(1+\ep)^3 .
\label{detb1}
\eeq
Subtraction of suitable non-negative terms from the right-hand side
of this inequality yields
\beq
\det (I+\ep^0) \ge (1-\ep)^3(1-30\ep) .
\label{detb2}
\eeq

Explicitly,
we have
\beq 
\ep^0=\ga^0 (c^{\mu 0}
{\ga}_{\mu}+d^{\mu 0}{\ga}_{5}
{\ga}_{\mu}+e^0+if^0{\ga}_{5}
+\frac{1}{2}g^{\la \mu 0}
{\sigma}_{\la \mu}).
\eeq
Noting the antisymmetry properties
of ${\sigma}_{\la \mu}$ and $g^{\la \mu \nu}$,
we see that $\ep^0$ is the sum of 16 terms,
each being a product of one Lorentz- and CPT-violating parameter
with one of the 16 gamma matrices.
Since the absolute value of an arbitrary entry
of any gamma matrix does not exceed 1,
it follows from the definition (\ref{defd}) of $\delta^0$ 
that $\ep \le 16\delta^0$.
Together with (\ref{detb2}),
this implies
\beq
\det (\ga^0 \Ga^0)>0,
\quad 0 \le \delta^0 < \frac{1}{480} .
\label{bounddet}
\eeq

In the trivial case $\delta^0=0$,
$\ga^0 \Ga^0=I$ has four positive eigenvalues.
The continuity of the determinant implies
this must also hold true for all $\delta^0$
in the above range.
An eigenvalue sign change would be accompanied
by a vanishing determinant,
contradicting (\ref{bounddet}).

\section{Bound for $\de$}

Equation \rf{disp} shows that 
the four roots of the dispersion relation
can be interpreted as eigenvalues of
$(\Ga^0)^{-1}(\Ga^j\la_j-M)$.
Note that the matrix $\Ga^0$ is invertible
provided the spinor redefinition (\ref{redef}) exists,
as we assume here.
We proceed by obtaining an upper bound on 
the quantity $\de$ in Eq.\ (\ref{defdel})
such that
\beq
\det(\ga^0\Ga^j\la_j -\ga^0M)\ne0 ,
\label{condition}
\eeq
where the factor of $\ga^0$ has been inserted for convenience.
With the bound on $\de$ in hand,
the continuity of the determinant in Eq.\ \rf{condition}
as the coefficients for Lorentz and CPT violation vanish 
then implies the same eigenenergy-sign structure 
as occurs in the usual Dirac case.

To simplify the notation,
define $\ep^j$ and $\ep(M)$
such that Eqs.\ \rf{Gam} and \rf{M}
take the forms
\beq
\Ga^j=\ga^j+\ga^0\ep^j ,
\quad
M=m+\ga^0\ep(M) .
\eeq
An argument similar to that following Eq.\ \rf{detb2}
shows the components $\ep_{kl}^j$ and $\ep_{kl}(M)$ 
of $\ep^j$ and $\ep(M)$ obey
\beq
m\ep_{kl}^j<16\delta,
\quad 
\ep_{kl}(M)<14\delta.
\label{boundM}
\eeq
Using this notation,
we can write
\beq
\ga^0(\Ga^j\la_j-M)
=\ga^0(\ga^j\la_j-m) +(\ep^j\la_j-\ep(M)) ,
\label{split}
\eeq
where the first term on the right-hand side is just
the usual free Dirac hamiltonian $H_D$
and the second term controls the Lorentz and CPT violation.

For the condition \rf{condition} to hold,
the kernel of $\ga^0(\Ga^j\la_j-M)$
must be empty. 
Thus,
$\ga^0(\Ga^j\la_j-M)v\ne0$
must hold for all complex spinors $v$.
The norm $|v|$ of $v$ can be set to 1 without
loss of generality.
A sufficient condition for the vanishing of the kernel
is then
\beq
|H_D v|^2 >|(\ep^j\la_j-\ep(M))v|^2 
\label{condit}
\eeq
for all $v$,
where we have used Eq.\ \rf{split}.

The left-hand side of this inequality is just $\vec{\la}^2+m^2$,
as can be seen by expanding $v$ in eigenspinors of $H_D$.
An upper bound for the right-hand side is determined by
$64(\sqrt{3}\cdot 8|\vec{\la}|+7m)^2 \delta^2$,
where we have used Eq.\ \rf{boundM}
and the assumption $|v|=1$.
Some algebra then directly yields the bound on $\de$ given in the text.

\end{appendix}

\end{multicols}

\begin{references}
\bibitem{fn0}
A third type of transformation
boosting and rotating both the particle fields 
and the coefficients within a fixed inertial frame can also be defined
\cite{ck}.
It is distinct from the others 
and is sometimes called an (inverse) active Lorentz transformation.
\bibitem{fn1}
One obstacle is that
the dispersion relation is a fourth-order polynomial in $\la^0$.
The expressions for the eigenenergies typically contain quartic roots 
involving $\vec{\la}$ and the coefficients for Lorentz violation.
A fourth-order term can therefore generate 
a first-order energy correction.
Another difficulty is that the momentum is a variable.
For example, 
even if the momentum component $p_z$ 
in the $z$ direction occurs only at second order,
when large it may dominate other first-order terms.
\bibitem{fn2}
Note that what we are calling the light cone may be unrelated 
to the properties of light in the present case.
In the standard-model extension,
the appropriate cone is defined by the behavior of observables 
under observer Lorentz transformations.
This is determined by the
spontaneous nature of the Lorentz violation,
which preserves the Minkowski metric
associated with the fundamental theory at the Planck scale. 
Nonetheless,
for simplicity we keep the terminology light cone. 
Note also that throughout this paper
the dispersion relations are in momentum space, 
so all references are to the energy-momentum light cone 
rather than the spacetime light cone.
\bibitem{fn3}
The coefficients $c_{\mu\nu}$ in the lagrangian \rf{lagr}
are defined to be traceless,
so a model with only a nonzero $c_{00}$
might appear inappropriate.
However,
the effect of a nonzero trace in $c_{\mu\nu}$
is merely to scale the normalization of the Dirac field
\cite{ck}.
For example,
a model with a nonzero $c_{00} = c$ describes physics similar 
to a model with $c_{00} = 3c/4$, $c_{jj} = c/4$, $j = 1,2,3$.
The former choice has been adopted in the text
to simplify the presentation.
\bibitem{fn4}
If $c_{00}$ is positive,
the issues with stability can be eliminated by multiplication 
with the exponential factor $\exp(-c_{00} \la_0^2)$ instead.
Thus,
consistency in the $c_{00}$ model can be achieved
through exponential suppression of this type
only for one sign of $c_{00}$ at a time.
Note that it is common to find field theories in which consistency
allows only a particular range for a parameter 
in the lagrangian.
For example,
stability requires that the squared-mass term 
in a scalar field theory have only one sign.
\end{references}
\end{document}